\documentclass[aip,apl,reprint,final]{revtex4-2}

\pdfoutput=1

\usepackage[colorlinks=true,allcolors=blue]{hyperref}
\usepackage{graphicx}% Include figure files
\usepackage{amsmath,amssymb,mathrsfs,array,longtable,delarray}
\usepackage{dcolumn}% Align table columns on decimal point
\usepackage{bm}% bold math

\begin{document}

\title{Stable electroluminescence in ambipolar dopant-free lateral p-n junctions}

\author{Lin Tian}
\affiliation{Institute for Quantum Computing, University of Waterloo, Waterloo N2L 3G1, Canada}
\affiliation{Department of Electrical and Computer Engineering, University of Waterloo, Waterloo N2L 3G1, Canada}
\author{Francois Sfigakis}
\altaffiliation{corresponding author: francois.sfigakis@uwaterloo.ca}
\affiliation{Institute for Quantum Computing, University of Waterloo, Waterloo N2L 3G1, Canada}
\affiliation{Department of Chemistry, University of Waterloo, Waterloo N2L 3G1, Canada}
\affiliation{Northern Quantum Lights Inc., Waterloo N2B 1N5, Canada}
\author{Arjun Shetty}
\affiliation{Institute for Quantum Computing, University of Waterloo, Waterloo N2L 3G1, Canada}
\affiliation{Department of Chemistry, University of Waterloo, Waterloo N2L 3G1, Canada}
\author{Ho-Sung Kim}
\affiliation{Department of Electrical and Computer Engineering, University of Waterloo, Waterloo N2L 3G1, Canada}
\affiliation{Waterloo Institute for Nanotechnology, University of Waterloo, Waterloo N2L 3G1, Canada}
\author{Nachiket Sherlekar}
\affiliation{Institute for Quantum Computing, University of Waterloo, Waterloo N2L 3G1, Canada}
\affiliation{Department of Physics and Astronomy, University of Waterloo, Waterloo N2L 3G1, Canada}
\author{\\Sara Hosseini}
\affiliation{Institute for Quantum Computing, University of Waterloo, Waterloo N2L 3G1, Canada}
\affiliation{Department of Electrical and Computer Engineering, University of Waterloo, Waterloo N2L 3G1, Canada}
\author{Man Chun Tam}
\affiliation{Department of Electrical and Computer Engineering, University of Waterloo, Waterloo N2L 3G1, Canada}
\affiliation{Waterloo Institute for Nanotechnology, University of Waterloo, Waterloo N2L 3G1, Canada}
\author{Brad van Kasteren}
\affiliation{Institute for Quantum Computing, University of Waterloo, Waterloo N2L 3G1, Canada}
\affiliation{Department of Electrical and Computer Engineering, University of Waterloo, Waterloo N2L 3G1, Canada}
\author{Brandon Buonacorsi}
\affiliation{Institute for Quantum Computing, University of Waterloo, Waterloo N2L 3G1, Canada}
\affiliation{Department of Physics and Astronomy, University of Waterloo, Waterloo N2L 3G1, Canada}
\author{Zach Merino}
\affiliation{Institute for Quantum Computing, University of Waterloo, Waterloo N2L 3G1, Canada}
\affiliation{Department of Physics and Astronomy, University of Waterloo, Waterloo N2L 3G1, Canada}
\author{\\Stephen R. Harrigan}
\affiliation{Institute for Quantum Computing, University of Waterloo, Waterloo N2L 3G1, Canada}
\affiliation{Department of Physics and Astronomy, University of Waterloo, Waterloo N2L 3G1, Canada}
\affiliation{Waterloo Institute for Nanotechnology, University of Waterloo, Waterloo N2L 3G1, Canada}
\author{Zbigniew Wasilewski}
\affiliation{Institute for Quantum Computing, University of Waterloo, Waterloo N2L 3G1, Canada}
\affiliation{Department of Electrical and Computer Engineering, University of Waterloo, Waterloo N2L 3G1, Canada}
\affiliation{Northern Quantum Lights Inc., Waterloo N2B 1N5, Canada}
\affiliation{Waterloo Institute for Nanotechnology, University of Waterloo, Waterloo N2L 3G1, Canada}
\affiliation{Department of Physics and Astronomy, University of Waterloo, Waterloo N2L 3G1, Canada}
\author{Jonathan Baugh}
\altaffiliation{baugh@uwaterloo.ca}
\affiliation{Institute for Quantum Computing, University of Waterloo, Waterloo N2L 3G1, Canada}
\affiliation{Department of Chemistry, University of Waterloo, Waterloo N2L 3G1, Canada}
\affiliation{Northern Quantum Lights Inc., Waterloo N2B 1N5, Canada}
\affiliation{Waterloo Institute for Nanotechnology, University of Waterloo, Waterloo N2L 3G1, Canada}
\affiliation{Department of Physics and Astronomy, University of Waterloo, Waterloo N2L 3G1, Canada}
\author{Michael E. Reimer}
\affiliation{Institute for Quantum Computing, University of Waterloo, Waterloo N2L 3G1, Canada}
\affiliation{Department of Electrical and Computer Engineering, University of Waterloo, Waterloo N2L 3G1, Canada}
\affiliation{Northern Quantum Lights Inc., Waterloo N2B 1N5, Canada}
\affiliation{Department of Physics and Astronomy, University of Waterloo, Waterloo N2L 3G1, Canada}

\begin{abstract}
Dopant-free lateral p-n junctions in the GaAs/AlGaAs material system have attracted interest due to their potential use in quantum optoelectronics (e.g., optical quantum computers or quantum repeaters) and ease of integration with other components, such as single electron pumps and spin qubits. A major obstacle to integration has been unwanted charge accumulation at the p\nobreakdash-n junction gap that suppresses light emission, either due to enhanced non-radiative recombination or inhibition of p-n current. Typically, samples must frequently be warmed to room temperature to dissipate this built-up charge and restore light emission in a subsequent cooldown. Here, we introduce a practical gate voltage protocol that clears this parasitic charge accumulation, in-situ at low temperature, enabling the indefinite cryogenic operation of devices. This reset protocol enabled the optical characterization of stable, bright, dopant-free lateral p-n junctions with electroluminescence linewidths among the narrowest ($<$ 1 meV; $<$ 0.5 nm) reported in this type of device. It also enabled the unambiguous identification of the ground state of neutral free excitons (heavy and light holes), as well as charged excitons (trions). The free exciton emission energies for both
photoluminescence and electroluminescence are found to be nearly identical (within 0.2 meV or 0.1 nm). The binding and dissociation energies for free and charged excitons are reported. A free exciton lifetime of 237 ps was measured by time-resolved electroluminescence, compared to 419 ps with time-resolved photoluminescence.
\end{abstract}

\maketitle

Due to their ease of integration with other optoelectronic devices, dopant-free lateral p-n junctions (2D planar light-emitting transistors) in GaAs/AlGaAs heterostructures,\cite{Dai13,Dai14,ChungYC19,HsiaoTK20} hosting two-dimensional electron gases (2DEGs) and hole gases (2DHGs), are a promising platform for photonic applications. An ultimate goal of quantum optoelectronics is the realization of an all-electrical, deterministic source of quantum states of light. By combining a source of single electrons (e.g., a non-adiabatic single electron pump) with a lateral p-n junction, the on-demand generation of single or entangled photons is possible.\cite{Blumenthal07,Brandon21,HsiaoTK20} Such sources would have immediate relevance for practical quantum sensing, communication and cryptography,\cite{Chunnilall14} and would benefit from the inherent scalability of lateral semiconductor devices (e.g., for multiplexing\cite{Laferriere20}). If such sources were integrated with spin qubits, spin-to-photon conversion schemes\cite{HsiaoTK20} could improve the scalability prospects for solid-state quantum computers. Another desirable application in quantum optoelectronics is photon-to-spin conversion,\cite{Oiwa17,Gaudreau17,FujitaT19} a necessary component of spin-based quantum repeaters for long-distance quantum communication.

Pioneering studies on lateral p-n junctions involved modulation-doped heterostructures and various selective etching techniques.\cite{Kaestner03A,Cecchini03,Cecchini04,Hosey04,Gell06} Early successes included observing the spin Hall effect through spin-to-photon conversion\cite{Wunderlich05} and demonstrating anti-bunching in the few-photon regime.\cite{Lunghi11} However, the presence of dopant impurities and an etched surface at the p-n junction itself can cause non-radiative recombination (reducing efficiency) and parasitic radiative recombination at different wavelengths.\cite{Kaestner03A,Kaestner03B}

\begin{figure*}[t]
    \includegraphics[width=1.9\columnwidth]{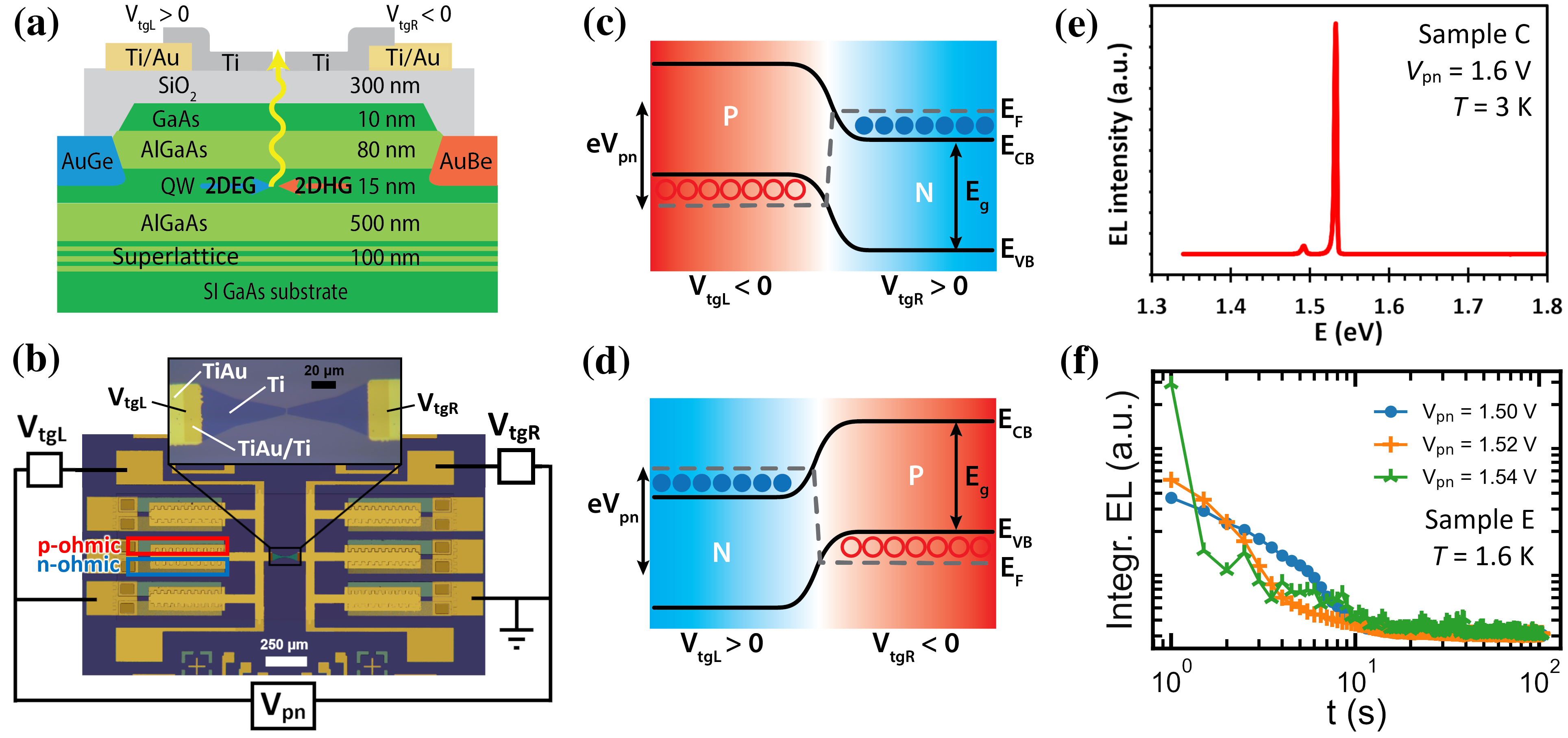}
    \caption{(a) Diagram of a dopant-free lateral p-n junction device. (b) Photograph of a completed device. $V_{\text{tgL}}$ is the topgate voltage on the left side, and $V_{\text{tgR}}$ is the topgate voltage on the right. $V_{\text{pn}}$ is the forward bias, used to drive current across the p-n junction. Note both sides of the p-n junction have ambipolar ohmic contacts, allowing a 2DEG or 2DHG to form on either side. Band structure schematic of a lateral p-n junction in: (c) the \textbf{PN mode} and (d) the \textbf{NP mode} (see text). Filled blue (empty red) circles represent electrons (holes) in a 2DEG (2DHG). (e) Typical EL spectrum. (f) Decaying EL of a device in DC mode (see text).}
    \label{Fig-FabDiode}
\end{figure*}

The limitations above can be circumvented by dopant-free field effect transistors.\cite{Harrell99,ChenJCH12,Croxall13,ChenJCH15,Deepyanti16} Furthermore, relative to their conventional modulation-doped counterparts, dopant-free devices have exceptional reproducibility and low disorder.\cite{Sarkozy09-A,Wendy10,See12,Wendy13,Sebastian16,Srinivasan20,Brandon21} Dopant-free p-n junctions have already been demonstrated,\cite{Dai13,Dai14,ChungYC19,HsiaoTK20} which generally show narrower electroluminescence (EL) emission peak linewidths than their modulation-doped counter-parts. A dopant-free single-photon source driven by surface acoustic waves was also recently realized.\cite{HsiaoTK20} However, there are reports, consistent with our experience (see Figure \ref{Fig-FabDiode}), that the brightness of dopant-free p\nobreakdash-n junctions can rapidly decay with time, requiring frequent thermal cycling from cryogenic to room temperature to restore (reset) the electrical properties and brightness of the device.\cite{HsiaoTK20,Dobney23}

In this Letter, we overcome the quenching of EL in dopant-free GaAs/AlGaAs quantum wells (QW) by implementing a sequence of gate voltages in-situ at low temperature to completely restore the device without the need for thermal cycling. We report on the narrowest EL linewidths observed to date in lateral p\nobreakdash-n junctions, whether doped or undoped. Well-defined EL emission peaks are visible up to a temperature of $T=85$~K, which we unambiguously identify as the ground state of neutral free excitons (labeled X$^0$ for heavy holes and LH for light holes) and the ground state of a heavy hole trion. In all samples, the free exciton emission energies for both
photoluminescence (PL) and EL are found to be nearly identical, with a symmetric lineshape. Using pulsed-EL, we report an exciton lifetime of 237~ps, much shorter than 419~ps obtained by pulsed-PL. The implemented gate voltage sequence, which we call the Set-Reset sequence, is a significant step towards realizing viable quantum light sources based on dopant-free 2DEGs and 2DHGs.\cite{Blumenthal07,Brandon21}

Data from five dopant-free lateral p\nobreakdash-n junctions (samples A, B, C, D, and E) is reported here, with a p\nobreakdash-n junction gap monotonically increasing from 200 nm (sample A) to 2000 nm (sample E). Samples A$-$D were fabricated on wafer G375, and sample E was fabricated on wafer G569. Both wafers contain a 15 nm wide GaAs QW and are otherwise identical except for the presence (absence) of a smoothing superlattice in G375 (G569) in the buffer layers. Typical mobilities for G375 and G569 are (3$-$6)$\times 10^{5}$ cm$^{2}$/Vs for 2DEGs and (1.7$-$3.0)$\times 10^{5}$ cm$^{2}$/Vs for 2DHGs from measurements on dedicated unipolar gated Hall bars (see Section X of the supplementary material for their complete characterization). Figure \ref{Fig-FabDiode} shows the growth, fabrication, and operation of dopant-free lateral p-n junctions (additional details are available in Sections I, II, and III of the supplementary material).  All EL spectra shown in this paper were taken with $|V_{\text{tgR}}|= |V_{\text{tgL}}|=5$~V.

A dopant-free lateral p-n junction is operated in \textbf{DC mode} when $V_{\text{pn}}$, $V_{\text{tgL}}$, and $V_{\text{tgR}}$ are kept at constant values. In addition, due to their ambipolar functionality, our p-n junctions can also be operated either in the \textbf{PN mode} [see Fig.~\ref{Fig-FabDiode}(c)] or in the \textbf{NP mode} [see Fig.~\ref{Fig-FabDiode}(d)]. Figure \ref{Fig-FabDiode}(e) shows a typical, wide-energy EL spectrum.

\begin{figure}[t]
    \includegraphics[width=1.0\columnwidth]{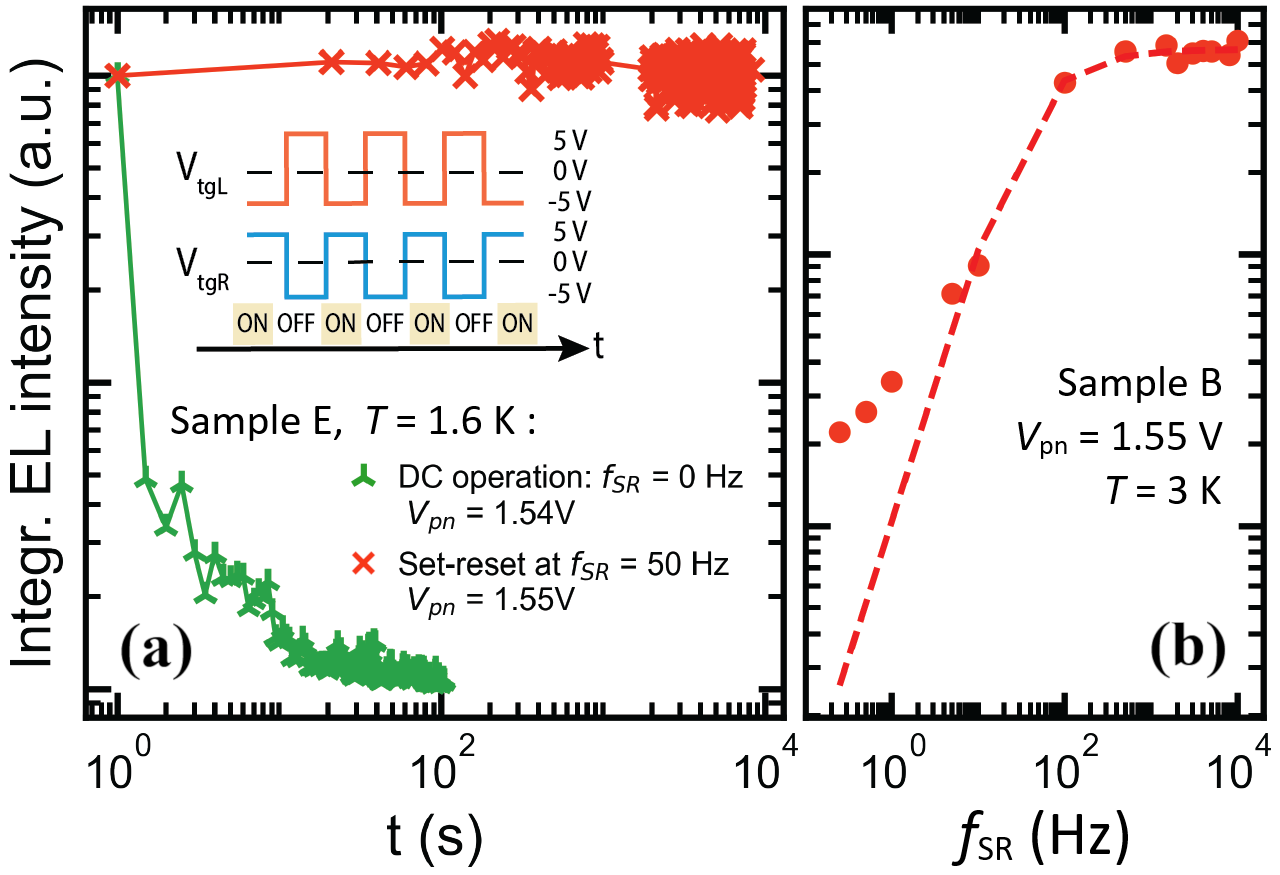}
    \caption{(a)(inset) Diagram of the Set-Reset voltage sequence for topgates with time, while $V_{\text{pn}}$ is held constant. (a)(main) EL in DC mode (green triangles) and in Set-Reset mode (red crosses). (b) EL (red circles) as a function of $f_{\textsc{sr}}$. The dashed line is a fit to Eqn.~(\ref{eq:intensity}).}
    \label{Fig-ACdriving}
\end{figure}

Figure \ref{Fig-FabDiode}(f) illustrates the principal problem in dopant-free lateral p\nobreakdash-n junctions: EL decays with time and vanishes within seconds. This quenching of EL is almost always accompanied by a similar quenching of the p\nobreakdash-n current. Significantly, with higher forward bias $V_{\text{pn}}$ (and hence with higher initial p\nobreakdash-n currents), light emission is suppressed more quickly. This is indicative of charging effects, either enhancing non-radiative electron-hole recombination or suppressing current altogether. Light emission can only be recovered in a subsequent cooldown if ``dark'' samples are warmed up to room temperature and electrically grounded for several hours.

As an alternative to a full thermal cycle, a Set-Reset voltage sequence applied to both $V_{\text{tgL}}$ and $V_{\text{tgR}}$ topgates can completely reset a dopant-free lateral p-n junction in-situ at low temperatures, such that light emission is fully recovered in the same cooldown. The Set-Reset sequence involves alternating the polarities of $V_{\text{tgL}}$ and $V_{\text{tgR}}$, while keeping $V_{\text{pn}}$ fixed (e.g., $V_{\text{pn}}=0$). The periodicity of the voltage sequence is determined by the Set-Reset frequency $f_{\textsc{sr}}$. In effect, the device alternates between PN mode and NP mode (with $V_{\text{pn}}=0$). For the reset to be most effective, (i) the magnitude of the topgate voltages must be large enough to alternately induce a 2DEG and 2DHG on each side of the p-n junction, and (ii) the voltage sequence must contain many cycles (50-500), where each cycle switches the topgate voltage polarities back and forth once. Alternating the polarity of $V_{\text{pn}}$ while holding $V_{\text{tgL}}$ and $V_{\text{tgR}}$ constant does not reset a device.

However, once reset, a device can still degrade again. Instead of applying the Set-Reset sequence before/after a set of measurements, one can continuously apply the Set-Reset voltage sequence with $V_{\text{pn}} > 1.52$ V (i.e., the bandgap of bulk GaAs), \textit{during} optical data acquisition. In this configuration, the Set-Reset sequence modulates the on/off states of the p\nobreakdash-n junction, by switching between forward bias (``set'') and reverse bias (``reset'') without changing $V_{\text{pn}}$ [see inset of Figure~\ref{Fig-ACdriving}(a)]. We call this operating regime the \textbf{Set-Reset mode}.

Figure \ref{Fig-ACdriving}(a) illustrates the dramatic difference between the DC mode and the Set-Reset mode. In DC mode, EL emission disappears very rapidly ($<$10 seconds). In stark contrast, the Set-Reset mode yields an EL intensity that does not decay over at least 10$^4$ seconds. In fact, it can remain bright for at least 48 hours, the longest period over which EL intensity was continuously tracked. Crucially, optical characteristics are reproducible for a given set of experimental parameters.

As $f_{\textsc{sr}}$ increases from 0.25 Hz to 500 Hz, Figure \ref{Fig-ACdriving}(b) shows EL becomes brighter: the average intensity during the light emitting portion (``set'') of a single set-reset cycle increases as the frequency becomes larger. Assuming emission intensity $I_\textsc{el}$ decays as $I_\textsc{el}(t) = I_0~e^{-t/\tau_d}$, where $I_0$ is the initial EL intensity at time $t=0$ and $\tau_d$ is the EL decay's mean lifetime (or half-life $\tau_d$\,ln\,2), the integrated intensity is:
\begin{equation}
I_\Sigma = \frac{\Delta t}{T_\textsc{sr}} \int_0^{T_\textsc{sr}} I_\textsc{el}(t)~dt,
\label{eq:integral}
\end{equation}
\noindent where $\Delta t$ is the data acquisition integration time, $T_\textsc{sr}$ is the period of a single Set-Reset cycle ($T_\textsc{sr} = 1/f_\textsc{sr}$), and the condition $T_\textsc{sr}<\Delta t$ is met. Performing the integral in Eqn.~(\ref{eq:integral}) on our ansatz for $I_\textsc{el}(t)$ yields:
\begin{equation}
I_\Sigma (f_\textsc{sr})= I_0\,\Delta t\,\tau_d\,f_\textsc{sr}\,(1-e^{-1/\tau_d f_\textsc{sr}}).
\label{eq:intensity}
\end{equation}
When $\tau_d f_\textsc{sr} \gg 1$, Eqn.~(\ref{eq:intensity}) predicts $I_\Sigma$ will saturate. In other words, when the set-reset period is very short ($T_\textsc{sr} \ll \tau_d$), the EL emission does not significantly decay during a single set-reset cycle, and $I_\Sigma$ becomes independent of $f_\textsc{sr}$. When the set-reset period is very long ($T_\textsc{sr} \gg \tau_d$), the EL emission decays significantly during a single set-reset cycle, and Eqn.~(\ref{eq:intensity}) predicts $I_\Sigma$ grows linearly with $f_\textsc{sr}$. The experimental data in Figure \ref{Fig-ACdriving}(b) is broadly consistent with Eqn.~(\ref{eq:intensity}) but, in the regime $T_\textsc{sr} > \tau_d$, the experimental $I_\Sigma$ is not a simple linear function of $f_\textsc{sr}$. Nevertheless, fitting Eqn.~(\ref{eq:intensity}) to this experimental data yields $\tau_d \approx 0.2$~seconds, which implies that the signal strength has already reduced by two orders of magnitude during the first second of data acquisition (from $t=0$ to $t=1$~s) when operating in DC mode.

\begin{figure}[t]
    \includegraphics[width=1.0\columnwidth]{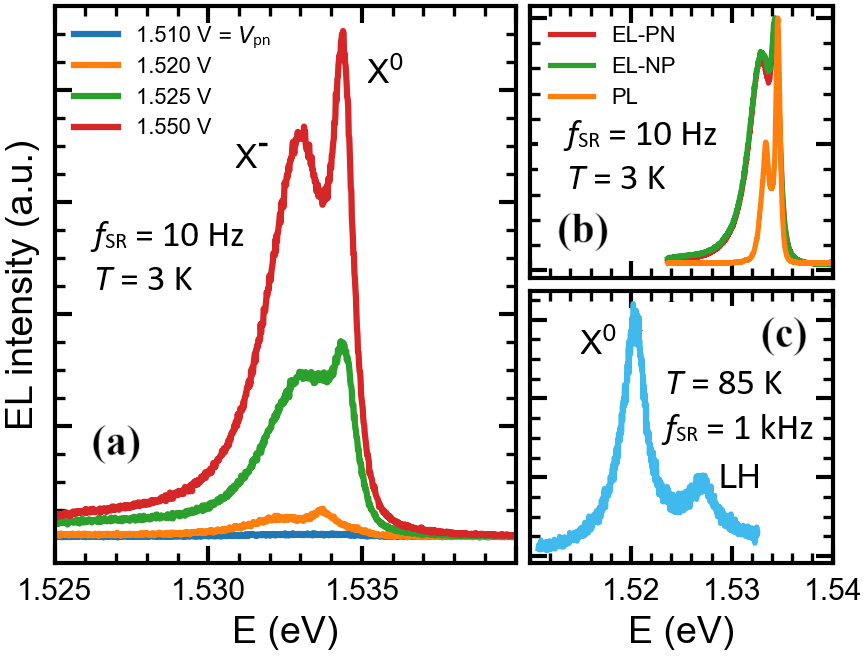}
    \caption{Sample C in the \textbf{Set-Reset mode}. (a) EL as a function of $V_{\text{pn}}$. (b) Comparison of EL spectra between PN mode and NP mode with $V_{\text{pn}} = 1.55$~V, and between PL and EL spectra. (c) Redshifted\cite{Passler01,Passler02} X$^0$ and LH peaks at $T=85$~K.}
    \label{Fig-Vpn-PNNP-ELPL-Tdep}
\end{figure}

\begin{table}[b]
    \begin{ruledtabular}
    \begin{tabular}{ccccccc}
    Sample & Gap & $E_{\textsc{el}}(\text{X}^0)$ & $E_{\textsc{el}}(\text{X}^-)$
    & $E_{\text{b}\textsc{x}}$ & $\Delta E_{\text{X}^-}$ & \textsc{fwhm} \\
    ~ & ($\mu$m) & (meV) & (meV) & (meV) & (meV) & (meV) \vspace{0.5mm}\\
    \hline
    A & 0.2 & 1534.1 & 1532.3 & 8.8 & 1.8 & 1.20 \\  % Sample A: 2DPN2-C4 (200 nm), was 9.1 meV
    B & 0.4 & 1534.3 & 1532.6 & 8.8 & 1.7 & 0.70 \\  % Sample B: 2DPN2-C2 (400 nm), was 8.9 meV
    C & 1.2 & 1534.4 & 1532.8 & 8.8 & 1.6 & 0.78 \\ % Sample C: 2DPN2-C3 (1200 nm), was 8.8 meV
    D & 1.2 & 1534.4 & 1532.8 & 8.8 & 1.6 & 0.92 \\ % Sample D: 2DPN2-C6 (1200 nm), was 8.8 meV
    E & 2.0 & 1534.7 & 1533.3 & 8.8 & 1.4 & 0.92  % Sample E: 2DPN7-C2 (2000 nm), was 8.5 meV
    \end{tabular}
    \end{ruledtabular}
    \caption{EL emission energies of X$^0$ and X$^-$ in all samples reported here. Also listed are their gaps between the \textit{p}-type and \textit{n}-type regions (i.e., the distance separating the $V_{\text{tgL}}$ and $V_{\text{tgR}}$ topgates in Figure \ref{Fig-FabDiode}(a)), their X$^0$ binding energies $E_{\text{b}\textsc{x}}$, their X$^-$ dissociation energies $\Delta E_{\text{X}^-} = E_{\textsc{el}}(\text{X}^0) - E_{\textsc{el}}(\text{X}^-)$, and the FWHM of their X$^0$ peaks.}
    \label{Table1}
\end{table}

Figure \ref{Fig-Vpn-PNNP-ELPL-Tdep}(a) shows EL spectra at different $V_{\text{pn}}$, with light emission occurring only once the forward bias exceeds the bandgap of bulk GaAs ($V_{\text{pn}}>1.519$~eV). Figure \ref{Fig-Vpn-PNNP-ELPL-Tdep}(b) shows EL spectra from the same p\nobreakdash-n junction in the PN and NP mode configurations.\footnote{Because of the continuous Set-Reset topgate voltage sequence, without physically changing any electrical connections, the ``ON'' state of the PN configuration corresponds to $V_{\text{pn}}=+1.55$~V, $V_{\text{tgL}}<0$, and $V_{\text{tgR}}>0$, whereas the ``ON'' state of the NP configuration corresponds to $V_{\text{pn}}=-1.55$~V, $V_{\text{tgL}}>0$, and $V_{\text{tgR}}<0$. Not all samples emit in PN and NP modes with equal EL intensities.} From their characteristic behavior in a detailed temperature dependence (not shown here), we can unambiguously attribute the narrowest peak ($E \sim 1.534$ eV) to the neutral exciton X$^0$ ground state for a 15 nm wide GaAs QW, and the lower-energy peak ($E \sim 1.533$ eV) to a negatively-charged exciton (trion) X$^-$.\cite{Bogardus68,Pavesi94} A complete justification for these two assignments can be found in Sections V and VIII of the supplementary material. At high temperatures, a new peak emerges [see ``LH'' in Figure \ref{Fig-Vpn-PNNP-ELPL-Tdep}(c)], which we unambiguously identify as
the ground state of light-hole free excitons [see Section IX in the supplementary material for justification].\cite{Shields95-B,DangGT09,OsborneJL96,Shields94,Oelgart94,OsborneJL96,Kundrotas08}
Of note, the PL and EL emission energies for X$^0$ match very well (within 0.2~meV or 0.1~nm). The emission energies of PL X$^0$ for both heterostructures G375 and G569 are nearly identical and consistent with literature (see Section VI in the supplementary material for the characterization of PL, including lineshape fits).\cite{Oelgart94,Birkedal96,OsborneJL96,Shields94,FinkelsteinG95,KumarR96,Manassen96,Hayne99,Esser00}

\begin{figure}[t]
    \includegraphics[width=1.0\columnwidth]{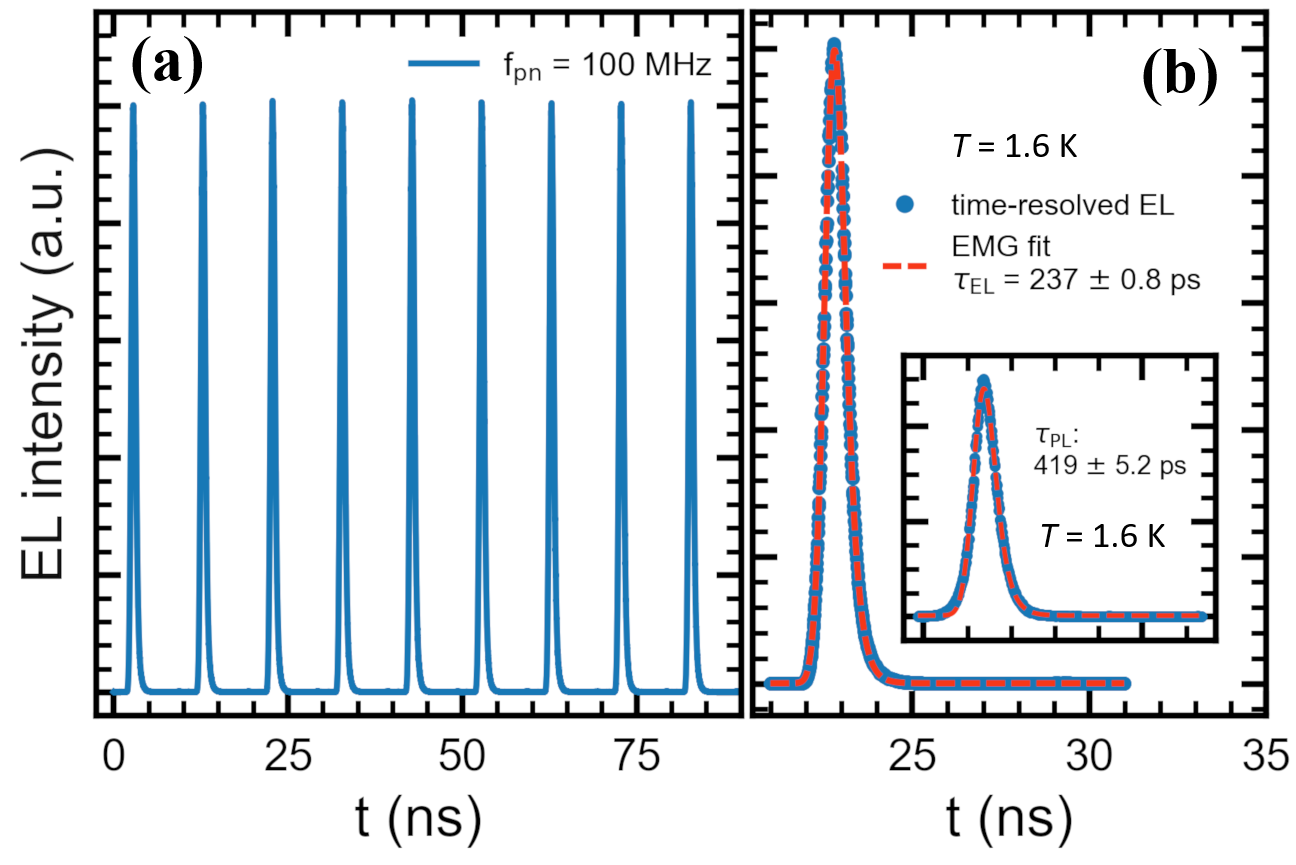}
    \caption{Time-resolved EL at $T=1.6$~K from sample E in Set-Reset mode with $f_{\textsc{sr}}=10$ Hz. (a) Periodic RF-pulsed EL with $V_{\text{pn}} = 1.47 \text{~V (dc)} + 0.5$~V$_{\text{p-p}}$ (\textsc{rf}). (b) Fitting the experimental data (blue circles) to an exponentially modified Gaussian (EMG; dashed red line), an EL exciton lifetime $\tau_\textsc{el}=237$ ps is obtained. (inset) The PL exciton lifetime from the same QW heterostructure is $\tau_\textsc{pl}=419$ ps. }
    \label{Fig-RF}
\end{figure}

\begin{figure*}[t]
    \includegraphics[width=2.0\columnwidth]{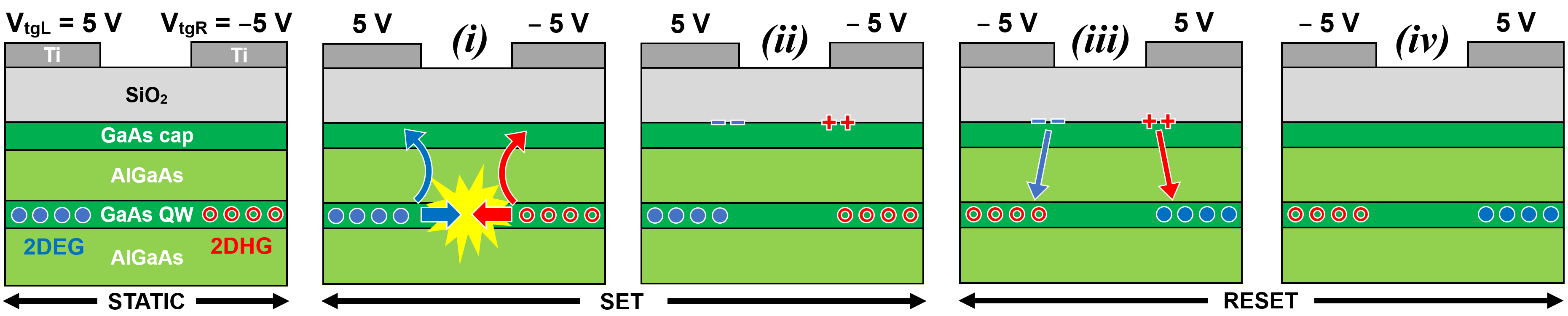}
    \caption{Diagram illustrating the parasitic charging and set-reset mechanisms. (\textsc{static}) The leftmost panel shows the device in DC mode with $V_{\text{pn}}=0$, and is the starting point of all optical experiments. This panel is not part of the Set-Reset sequence. The Set-Reset sequence is shown in panels \textit{(i)} through \textit{(iv)}, with $V_{\text{pn}}=+1.5$ V remaining constant in all four panels. (\textsc{set}) The next two panels show events during the ``set'' cycle of the Set-Reset sequence; the p-n junction is in the \textit{forward-bias} regime. \textit{(i)} Radiative electron-hole recombination with finite p-n current occurs (electroluminescence). A small proportion of electrons and holes escape the quantum well (QW) confinement, and head towards the GaAs/SiO$_2$ interface where they get trapped. \textit{(ii)} As the parasitic charges trapped at the GaAs/SiO$_2$ interface build up, EL decays. Eventually, the p-n current and light emission vanish altogether. (\textsc{reset}) The two rightmost panels show events during the ``reset'' cycle of the Set-Reset sequence; the p-n junction is in the \textit{reverse-bias} regime. \textit{(iii)} The polarity of both topgates has been reversed. The metastable electron (hole) charges trapped at the GaAs/SiO$_2$ interface are dislodged, repelled by the topgate above and attracted to the 2DHG (2DEG) below. \textit{(iv)} All previously built-up parasitic charges have now been drained to the 2DEG/2DHG nearby, and the p-n junction has been restored to its original state. The ``reset'' cycle ends, and the next ``set'' cycle can start [panel $(i)$ again].}
    \label{Fig-SetReset}
\end{figure*}

Table \ref{Table1} lists the emission energies and full width at half maximum (FWHM) of X$^0$ in all samples reported here, obtained by fitting the EL lineshape. Figure S4 in the supplementary material shows data and details about the EL lineshape fitting\cite{Stancik08,Manassen96} for samples A, B, C, and D. Four of the five samples listed in Table \ref{Table1} have narrower linewidths (0.7$-$0.9 meV) than the narrowest linewidths (1.0$-$1.6 meV) of any lateral p-n junctions reported in the literature, whether undoped\cite{Dai13,Dai14,ChungYC19,HsiaoTK20} or modulation-doped.\cite{Kaestner03A,Cecchini03,Cecchini04,Hosey04,Gell06} The most likely reason for such narrow linewidths is the high quality MBE growth, as evidenced by the clean PL spectrum shown in Figure S6 and the high electron/hole mobilities shown in Figure S10 of the supplementary material. However, we speculate that, in the vicinity of the p\nobreakdash-n junction, the Set-Reset voltage sequence clears away parasitic charge that causes additional scattering, and hence reduces the broadening of EL emission.

As the p-n junction gap between the $V_{\text{tgL}}$ and $V_{\text{tgR}}$ topgates decreases from 2000 nm to 200 nm, both X$^-$ and X$^0$ show a very weak Stark shift (0.6~meV) to lower EL energies in Table~\ref{Table1} due to the increasing in-plane electric field $|\vec{E}_{\text{ext}}| = V_{\text{pn}}$/gap in the 2DEG/2DHG plane. Calculations and assumptions used for determining exciton binding energies $E_{\text{b}\textsc{x}}$ are described in Section VII of the supplementary material.

The p-n junction gap length did not otherwise appear to have any other effect on transport or optical properties. The EL of all five samples quenched within mere seconds of starting light emission. The Set-Reset mode appears to have been equally effective at restoring light emission in all five samples.

Figure \ref{Fig-RF} demonstrates that the Set-Reset mode is compatible with radio frequency (RF) operation of lateral p\nobreakdash-n junctions.\footnote{The electrical circuit used is shown in Figure S2 from the supplemental material. The lineshape of the EL peaks in Figure \ref{Fig-RF} is dictated in part by the limitations of the RF equipment (max. rise time of 0.95 ns/0.6 V$_{\text{p-p}}$); the attenuated RF pulses reaching the samples are very unlikely to be square-shaped pulses. Nevertheless, the lateral p-n junction is clearly responsive on timescales of less than 1 ns [see Fig.\,\ref{Fig-RF}(a)].} The shorter lifetime of EL relative to PL is consistent with the wider FWHM observed in EL relative to PL [see Fig.\,\ref{Fig-Vpn-PNNP-ELPL-Tdep}(b)]. Hypothetically, in the single photon regime, an exciton lifetime $\tau_\textsc{el}=237$ ps would be compatible with a 1~GHz emission rate for the single photon source proposed in Refs.~\onlinecite{Blumenthal07,Brandon21}.

Five key observations support the scenario of localized parasitic charging in lateral p\nobreakdash-n junctions:\\
\noindent $(1)$ quenching of EL with time [see Figure \ref{Fig-FabDiode}(f)];\\
\noindent $(2)$ faster quenching of EL with larger initial forward \\ \indent ~~bias currents [see Figure \ref{Fig-FabDiode}(f)];\\
\noindent $(3)$ brighter EL when operating a device in the Set-Reset \\ \indent ~~mode [see Figure \ref{Fig-ACdriving}(a)];\\
\noindent $(4)$ the ``reset'' requiring the alternating presence of \\ \indent ~~both 2DEG and 2DHG in the same location to \\ \indent ~~be most effective (see below); and\\
\noindent $(5)$ the ``reset'' not requiring a finite $V_{\text{pn}}$ to be effective.

\textit{Regarding observations $(1)-(3)$.} Without current flowing across the p\nobreakdash-n junction, the 2DEG and 2DHG carrier densities are otherwise stable, before or after electroluminescence is quenched. This suggests that the charging mechanism making devices unstable is only associated with current flowing across the p\nobreakdash-n junction,\footnote{From our own experience and that of others,\cite{ShengDiPC2020} at very high p-n currents (0.2$-$0.4 mA), dopant-free p-n junctions can be stable in time. In that case, we speculate that any parasitic charge build-up is cleared away by the high currents. We believe this high-current regime is not applicable to the single photon regime, where currents are expected to be six orders of magnitude smaller.} and is localized near the p\nobreakdash-n junction since the 2DEG/2DHG themselves are not suppressed. We note that electrons/holes can escape the quantum well confinement in significant numbers at/near the p\nobreakdash-n junction when the forward bias provides energies ($eV_{\text{pn}}>1.5$~eV) much larger than the QW confinement potential, provided by the GaAs/AlGaAs conduction band offset for electrons ($\sim\,$0.217~eV) or GaAs/AlGaAs valence offset for holes ($\sim\,$0.164~eV). Another possible escape mechanism could be EL self-illumination: electrons (holes) from the QW are photoexcited into the GaAs cap layer by emitted EL photons, since the EL photon energy (1.534~eV) exceeds the AlGaAs barrier heights (band offsets).

\textit{Regarding observation $(4)$.} If the Set-Reset sequence is only applied to one side (say, $V_{\text{tgR}}$ but not $V_{\text{tgL}}$) of an ambipolar p\nobreakdash-n junction, then light emission lasts longer than in DC mode but not as long as when the Set-Reset sequence is applied to both $V_{\text{tgR}}$ and $V_{\text{tgL}}$. In addition to the five ambipolar p\nobreakdash-n junctions reported here, ``unipolar'' p\nobreakdash-n junctions were also fabricated, with only \textit{n}\nobreakdash-type ohmic contacts on one side of the p\nobreakdash-n junction and only \textit{p}\nobreakdash-type ohmic contacts on the other side. These were also unstable with time (i.e., EL quenching), similar to their ambipolar cousins. However, the Set-Reset sequences failed to reset these unipolar devices: only full thermal cycles could reset them. These two results together strongly suggest an efficient ``reset'' mechanism must involve the presence of alternating 2DEGs and 2DHGs in the same physical location, with its associated reversal of the electric field direction.

\textit{Regarding observation $(5)$.} The ``reset'' is effective whether $V_{\text{pn}}=0$ or $V_{\text{pn}}\neq 0$. Thus, unlike the mechanism behind parasitic charging, the ``reset'' mechanism does not involve any phenomena associated with $V_{\text{pn}}\neq 0$.

A scenario for parasitic local charging in dopant-free lateral p\nobreakdash-n junctions is illustrated in Figure \ref{Fig-SetReset}. During EL emission, driven by photoexcitation and/or the electric fields from $V_{\text{pn}}$, $V_{\text{tgL}}$, and $V_{\text{tgR}}$ present at the p\nobreakdash-n junction, electrons (holes) tunnel/escape from their 2DEG (2DHG) QW into the surrounding GaAs/AlGaAs material and to the GaAs/SiO$_2$ interface at the wafer surface. This parasitic charge build-up either enhances non-radiative electron-hole recombination, or counters the forward bias enough to altogether suppress current across the p\nobreakdash-n junction. By reversing the directions of the electric fields stemming from the topgates during the ``reset''cycle of the Set-Reset sequence, all or most of the trapped electron (hole) charges at the GaAs/SiO$_2$ interface are dislodged from their metastable traps and ``push-pulled'' to recombine with the newly-formed 2DHG (2DEG) below. We note the two tunneling processes in the ``Set'' and ``Reset'' cycles are not complementary/reversed processes: one requires $V_{\text{pn}}>1.5$~V (i.e., photoexcitation) to occur while the other does not (it can occur at $V_{\text{pn}}=0$).

In conclusion, we have proposed a mechanism (localized parasitic charging) for quenched electroluminescence in lateral p\nobreakdash-n junctions, and demonstrated an operating regime (the Set-Reset mode) that dissipates this parasitic charge. The Set-Reset mode allowed
the observation of the narrowest EL linewidths (0.70~meV) achieved to date in lateral p\nobreakdash-n junctions, and the indefinite operation of ambipolar lateral p\nobreakdash-n junctions at cryogenic temperatures (up to 85~K), obviating the need for frequent thermal cycles to room temperature. In turn, this enabled the unambiguous identification of the heavy-hole free exciton X$^0$, the light-hole free exciton LH, and charged exciton X$^-$. The emission energies for free excitons EL X$^0$ and PL X$^0$ were closely matched (within 0.2~meV or 0.1~nm) and both their lineshape were symmetric in all observed samples. Finally, we demonstrated the Set-Reset mode is compatible with RF (time-resolved) operation, a necessary condition for the future realization of single photon sources based on dopant-free lateral p-n junctions.

\section*{Supplementary Material}

The supplementary material provides additional details on MBE growth, sample fabrication, optics setups, experimental PL and EL spectra, lineshape fitting of EL and PL spectra, calculations for binding energies of excitons and trions, mobility and carrier density measurements, and band structure diagrams and simulations (using a self-consistent Poisson-Schr$\ddot{\mbox{o}}$dinger solver\cite{NextNano-A, NextNano-B, NextNano-C}).

\section*{Authors' Contribution}

The first three authors contributed equally to this work, and the last two authors contributed equally to this work.

\section*{Acknowledgments}

The authors thank Christine Nicoll for useful discussions. This research was undertaken thanks in part to funding from the Canada First Research Excellence Fund (Transformative Quantum Technologies), Defence Research and Development Canada (DRDC), and Canada's Natural Sciences and Engineering Research Council (NSERC). The University of Waterloo's QNFCF facility was used for this work. This infrastructure would not be possible without the significant contributions of CFREF-TQT, CFI, ISED, the Ontario Ministry of Research and Innovation, and Mike and Ophelia Lazaridis. Their support is gratefully acknowledged.

\section*{Data availability}

The data that support the findings of this study are available from the corresponding author upon reasonable request.

%aipnum4-2.bst 2019-01-14 (MD) hand-edited version of apsrev4-1.bst

\end{document}

\title{{\Large{SUPPLEMENTARY MATERIAL}}\\~\\
Stable electroluminescence in ambipolar dopant-free lateral p-n junctions}

\author{L. Tian}
\author{F. Sfigakis}
\author{A. Shetty}
\author{Ho.-S. Kim}
\author{N. Sherlekar}
\author{S. Hosseini}
\author{\\M. C. Tam}
\author{B. van~Kasteren}
\author{B. Buonacorsi}
\author{Z. Merino}
\author{S. R. Harrigan}
\author{\\Z. Wasilewski}
\author{J. Baugh}
\author{M. E. Reimer}

\affiliation{University of Waterloo, Waterloo N2L 3G1, Canada}

\begin{abstract}
~\\~\\
\indent\qquad\qquad\qquad\qquad \textbf{Table of Contents}\\~\vspace{-7mm}\\
Section \ref{sec:mbe}: Experimental methods, Molecular beam epitaxy \\
Section \ref{sec:fab}: Experimental methods, Sample fabrication \\
Section \ref{sec:optics}: Experimental methods, Optics \\
Section \ref{sec:diodes}: Ambipolar diode I-V traces \\
Section \ref{sec:ELfits}: Electroluminescence lineshape fits and peak identification \\
Section \ref{sec:PLfits}: Non-resonant photoluminescence of wafers G375 and G569 \\
Section \ref{sec:E_bX}: Binding energies of EL X$^0$ and EL X$^-$ \\
Section \ref{sec:Tdep}: Detailed temperature dependence of electroluminescence \\
Section \ref{sec:LH}: Identifying light holes in electroluminescence \\
Section \ref{sec:transport}: Transport characterization, mobility and carrier density\\
Section \ref{sec:nextnano}: Self-consistent band structure simulations
\end{abstract}

\maketitle

\section{Experimental methods, molecular beam epitaxy (MBE)}
\label{sec:mbe}

\begin{figure}[t]
    \includegraphics[width=0.95\columnwidth]{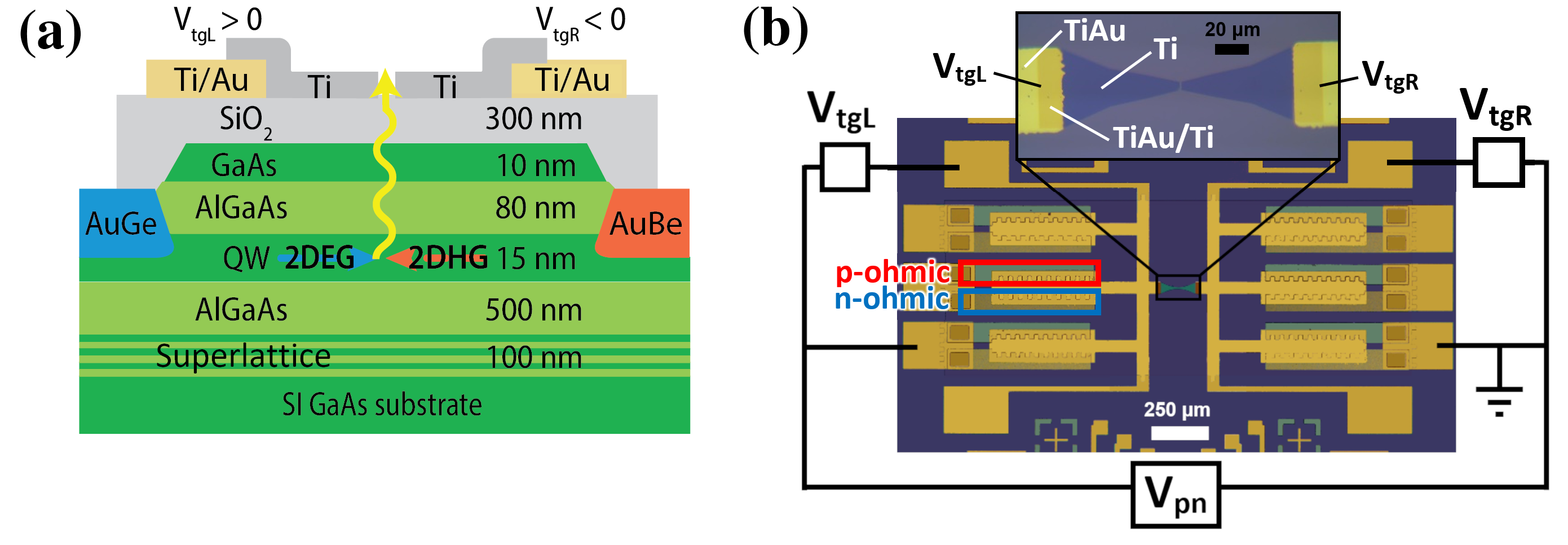}
    \caption{For convenience, Figures 1a and 1b from the main text are reproduced here. (a) Schematic diagram of a dopant-free lateral p-n junction device, including the epitaxial layers of the quantum well (QW) heterostructure G375. (b) Composite photograph of a dopant-free lateral p-n junction, as well as the electrical circuit used in all low-frequency measurements. Note both sides of the p-n junction have ambipolar ohmic contacts, allowing a two-dimensional electron gas (2DEG) or a two-dimensional hole gas (2DHG) to form on either side. $V_{\text{tgL}}$ is the topgate voltage on the left side, and $V_{\text{tgR}}$ is the topgate voltage on the right. $V_{\text{pn}}$ is the forward bias, used to drive current across the p-n junction. (\textsc{inset}) The black labels identify the left and right topgates in the vicinity of the p-n junction, which is itself located at the center of the photograph. The white labels identify the two layers (opaque TiAu, and semi-transparent Ti) composing the left/right topgates. Their overlap region is labeled TiAu/Ti.}
    \label{Fig-MBE-Fab}
\end{figure}

Two quantum well heterostructures, wafers G375 and G569, were grown by MBE on semi-insulating (SI) GaAs (100) substrates. Wafer G375 has the following layer sequence (starting from the substrate): a 200 nm GaAs buffer, a 20-period smoothing superlattice (SL) composed of a 2.5 nm GaAs layer and 2.5 nm Al$_{0.3}$Ga$_{0.7}$As layer, a 500 nm Al$_{0.3}$Ga$_{0.7}$As bottom barrier, a 15 nm wide GaAs quantum well, a 80 nm Al$_{0.3}$Ga$_{0.7}$As top barrier, and a 10 nm GaAs cap layer. There was no intentional doping anywhere in the heterostructure. Wafer G569 is nominally identical to G375 except that the 20-period smoothing superlattice buffer was replaced by a 500 nm LT-GaAs buffer followed by a 1,000 nm GaAs buffer. Wafer G375 (G569) was used to fabricate samples A$-$D (E).

\section{Experimental methods, Sample fabrication}
\label{sec:fab}

Hall bars and p-n  junctions were fabricated, all oriented in the high mobility crystal direction $[1\bar{1}0]$. The fabrication of unipolar/ambipolar Hall bars on dopant-free wafers is described in Refs.~\onlinecite{Wendy10,ChenJCH12,Deepyanti16}. Briefly, after a mesa etch, Ni/AuGe/Ni \textit{n}-type recessed ohmic contacts and AuBe \textit{p}-type recessed ohmic contacts were deposited and annealed at 450$^\circ$C for 180 seconds and at 520$^\circ$C for 180 seconds, respectively. A 300 nm thick SiO$_2$ insulator layer was deposited by plasma-enhanced chemical vapor deposition (PECVD). Above the SiO$_2$ insulator layer, a TiAu topgate covers the entire surface of the 2DEG or 2DHG (overlapping the ohmic contacts). Unlike conventional modulation-doped GaAs/AlGaAs heterostructures, it is this topgate that determines the shape of the 2DEG/2DHG, not the etched mesa. Near the p-n  junction, the topgate is composed of a single 5 nm thin semi-transparent Ti layer ($\sim$\,70\% transmission) in samples A$-$D and composed of a single 30 nm thin semi-transparent indium-tin-oxide (ITO) layer ($\sim$\,85\% transmission) for sample E. The topgate gap is the distance between the left topgate ($V_{\text{tgL}}$) and right topgate ($V_{\text{tgR}}$), and is also the distance between the \textit{p}-type and \textit{n}-type regions. The topgate gap varies from 200 nm to 2,000 nm in samples A$-$E (see Table 1 in the main text). This type of device architecture (\textit{n}\nobreakdash-type or \textit{p}\nobreakdash-type) is normally switched off unless a gate voltage beyond a turn-on threshold is applied, inducing a 2DEG (2DHG) underneath the gate with a positive (negative) voltage. Note the presence of \textit{n}-type and \textit{p}-type ohmic contacts on \text{both} sides of the our p-n  junctions, allowing a two-dimensional electron gas (2DEG) or a two-dimensional hole gas (2DHG) to form. The device can thus be operated in the PN, NP, NN, or PP modes. Electroluminescence occurs only in the PN or NP modes.

\begin{figure}[h]
    \includegraphics[width=0.65\columnwidth]{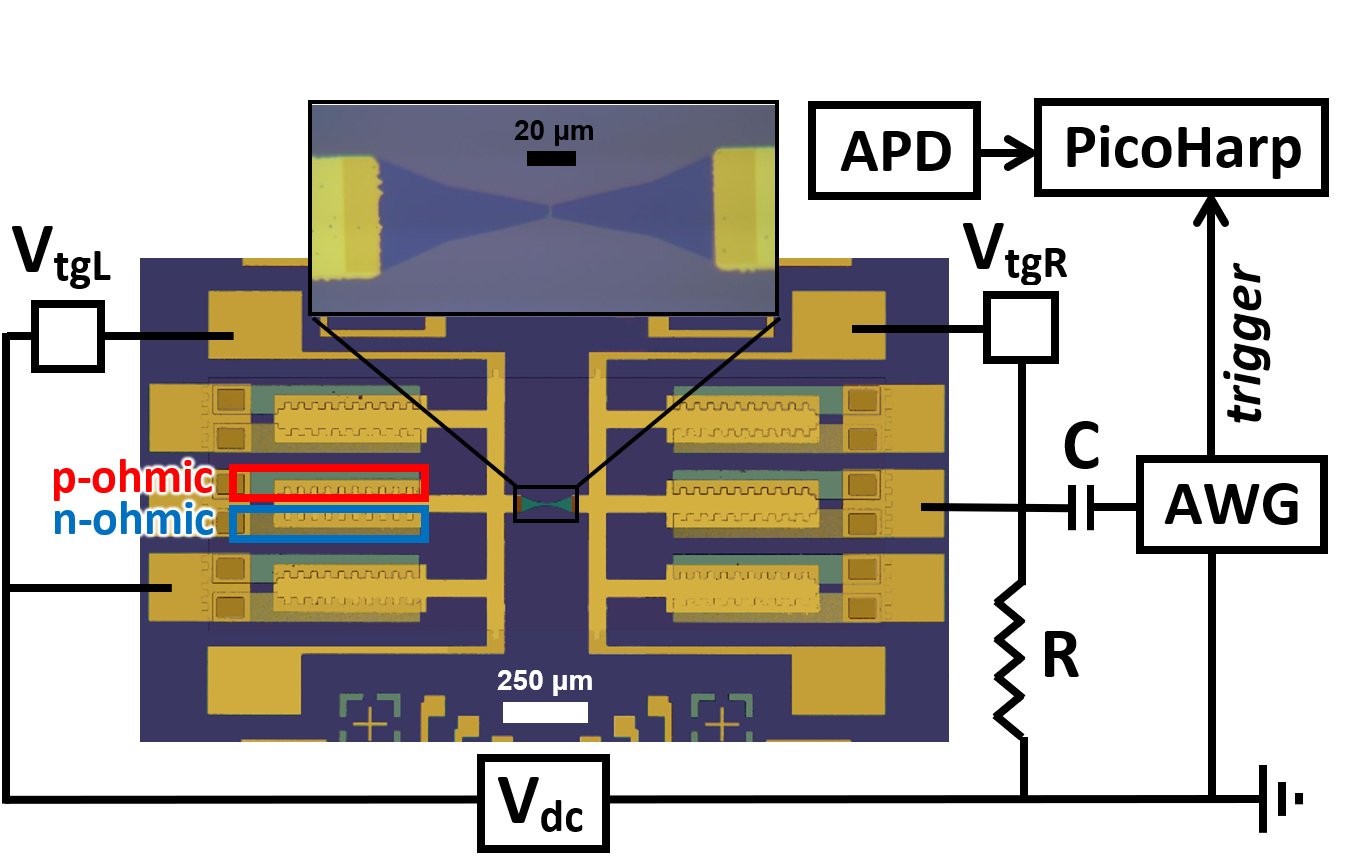}
    \caption{\textbf{Electrical RF circuit used for all time-resolved electroluminescence}.
     Note the forward bias is now $V_{\text{pn}} = V_{\text{dc}} + V_{\textsc{rf}}$, the latter being supplied by an arbitrary waveform generator (AWG) Tektronix 5014B. The RF pulse was a square wave with repeat frequency 100 MHz (10 ns period) with a nominal 10\% duty cycle.
     However, the AWG has a max. rise time of 0.95 ns/0.6 V$_{\text{p-p}}$, and outputs RF pulses that are Gaussian-shaped. A PicoHarp 200 provided time tagging using signals from the AWG and the avalanche photodiode (APD). The AWG generated a square wave with repeat frequency 100 MHz (10 ns period) with a nominal 10\% duty cycle. Periodic RF-pulsed EL was obtained with $V_{\text{pn}} = 1.47 \text{~V (dc)} + 0.5$~V$_{\text{p-p}}$ (\textsc{rf}).}
    \label{FigS-RFcircuit}
\end{figure}

\section{Experimental methods, Optics}
\label{sec:optics}

All spectra were acquired with a spectrometer grating with a groove density of 1,800 lines/mm, except for Figure 1(e) in the main text (150 lines/mm). Devices were cooled down either to $T=3$~K in an Oxford Instruments OptistatDry BLV closed-cycle cryostat or to $T=1.6$~K in an Attocube AttoDRY2100 closed-cycle cryostat, both with in-house customized DC and RF electrical feedthroughs. Unless specified otherwise, cw photoluminescence was generated from photoexcitation with a HeNe laser (632.8 nm) at 1 mW optical output power. Pulsed photoexcitation for time-resolved photoluminescence was provided by a Ti:sapphire laser operating at 800 nm. The electrical circuit and optical setup for time-resolved electroluminescence is shown in Figure \ref{FigS-RFcircuit}.

\section{Ambipolar diode I-V traces}
\label{sec:diodes}

\begin{figure}[h]
    \includegraphics[width=1.0\columnwidth]{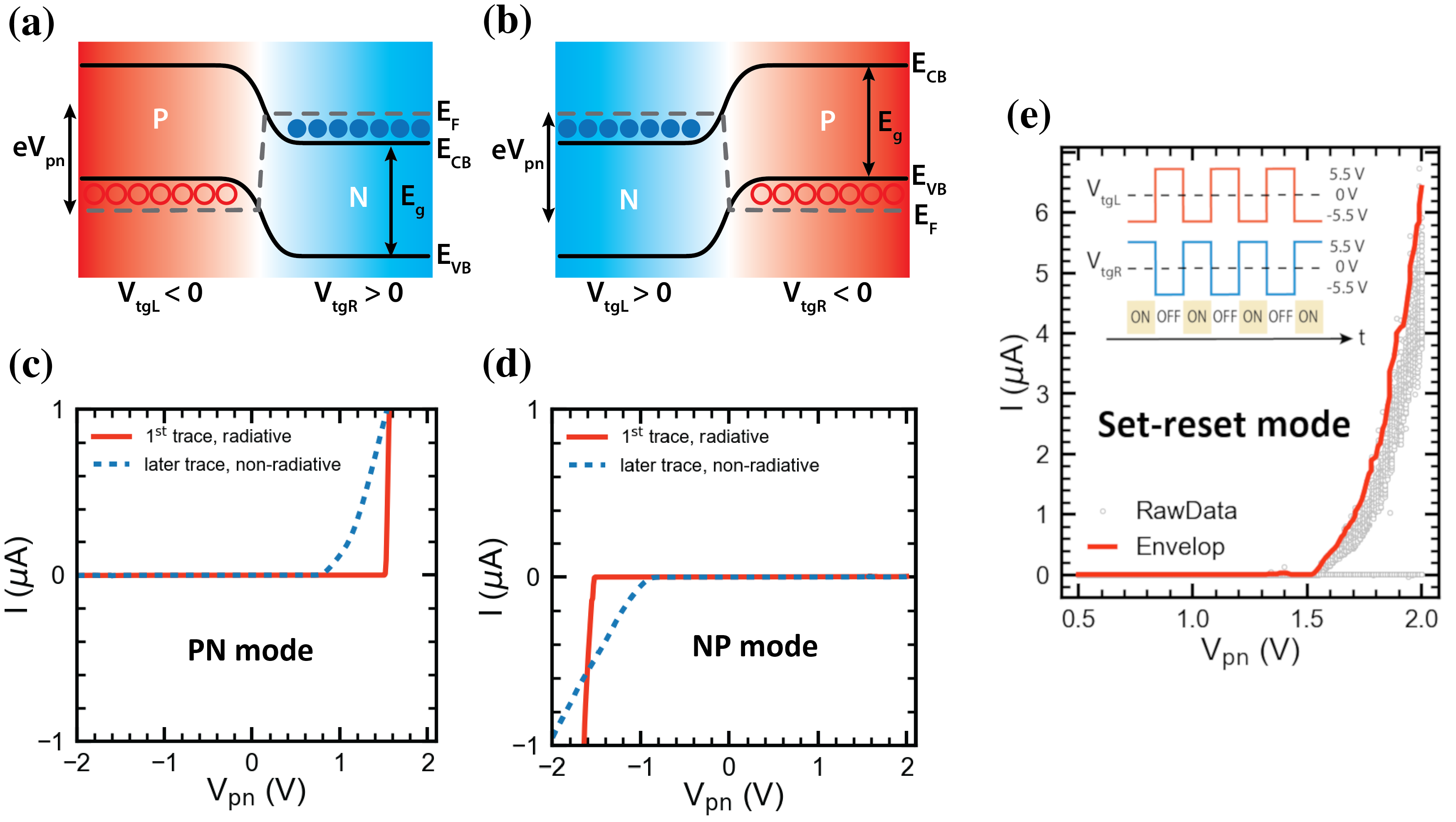}
    \caption{Band structure schematic across the p-n junction in the quantum well plane, when the device is operated in: (a) \textbf{PN mode} and (b) \textbf{NP mode}; filled blue circles represent electrons in a 2DEG, and empty red circles represent holes in a 2DHG. Diode current when sample C is operated in: (c) PN mode with $V_{\text{tgL}} = -5$~V and  $V_{\text{tgR}} = +5$~V, and (d) NP mode with $V_{\text{tgL}} = +5$~V and  $V_{\text{tgR}} = -5$~V. (e)(\textsc{inset}) Diagram of the Set-Reset voltage sequence for topgates with time, while $V_{\text{pn}}$ is held constant. If $V_{\text{pn}} > +1.5$ V, then the diode switches between ON/OFF states, as indicated by the highlighted labels. (e)(\textsc{main}) Raw diode I-V curve in \textbf{Set-Reset mode.} The grey circles are the raw data, twenty data points for each value of forward bias $V_{\text{pn}}$. Current is measured for both ON and OFF cycles. The red line is the highest current at a given $V_{\text{pn}}$ (the ``envelop'' of the dataset). Our instrument undersampled the data: measurements were performed below the Nyquist frequency (with respect to the set-reset frequency $f$), and suffered from a long RC time constant (exceeding $1/f$). We believe that even the envelop current is an under-estimate of the true p-n current.}
    \label{Fig-Diode}
\end{figure}

Figure \ref{Fig-Diode}a shows the schematic band structure of a dopant-free lateral p-n junction operated in \textbf{PN mode}, defined as when $V_{\text{tgL}}<0$ and  $V_{\text{tgR}}>0$, whereas Figure \ref{Fig-Diode}b shows the band structure of the same device operated in \textbf{NP mode}, when $V_{\text{tgL}}>0$ and  $V_{\text{tgR}}<0$. Figure \ref{Fig-Diode}c demonstrates the measured diode behavior when the lateral p-n junction is operated in PN mode (red solid trace). Current only flows when $V_{\text{pn}} > +1.5$~V in forward bias, and no current flows in reverse bias (i.e., $V_{\text{pn}} < 0$). The diode turn-on threshold is at the expected value for the GaAs bandgap ($\sim$\,1.5~eV).  The ambipolar device behaves identically when operated in the NP mode (see Figure \ref{Fig-Diode}d). However, all samples showed strong hysteretic behavior after the first $V_{\text{pn}}$ sweep: each subsequent sweep had a different turn-on $V_{\text{pn}}$ bias and ideality factor. The blue dashed lines in Figures \ref{Fig-Diode}c,d are examples of such non-reproducible sweeps. Most importantly, radiative electron-hole recombination only occurred during the first $V_{\text{pn}}$ sweep. Subsequent $V_{\text{pn}}$ sweeps usually did not produce any light emission, unless either $V_{\text{pn}}$ or $V_{\text{tg}}$ (we use $V_{\text{tg}}$ to refer to either/both $V_{\text{tgR}}$ or $V_{\text{tgL}}$.) were increased beyond values used in previous sweeps. After a thermal cycle to room temperature and back down to cryogenic temperatures, device characteristics are fully restored: red trace lines in Figs. \ref{Fig-Diode}c and \ref{Fig-Diode}d, with its associated light emission. Finally, Figure \ref{Fig-Diode}e shows a typical diode I-V trace when a devices is operated in the \textbf{Set-Reset} mode. The measurement in this mode is very noisy because 2DEGs and 2DHGs continuously form and dissolve on both sides of the p-n junctions, causing currents to flow in/out to (de)populate the 2DEG/2DHG. Overall however, the diode turn-on remains at the correct threshold, $\sim$1.5 V for GaAs.

\clearpage

\begin{figure}[t]
    \includegraphics[width=0.95\columnwidth]{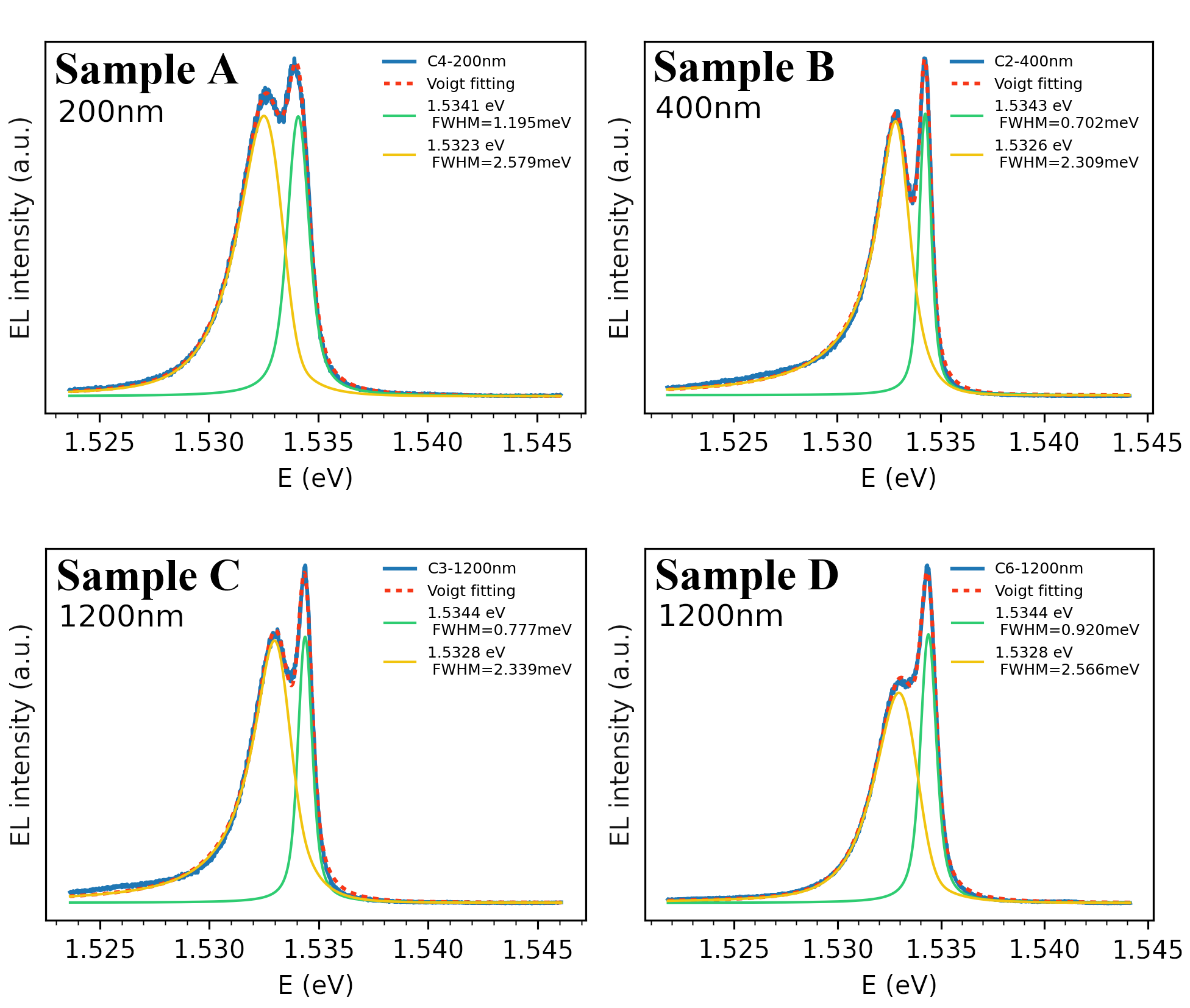}
    \caption{EL spectra at $T=3$~K for samples A$-$D. The four panels above show the lineshape fits for the electroluminescence spectrum of each sample in Table I, with the resulting fitting parameters listed in each panel for both the neutral free exciton and trion peaks. The solid blue line is the experimental data. The dashed red line is the integrated fit, including contribution from both peaks. Symmetric pseudo-Voigt functions\cite{Stancik08} (green solid lines) were used for fitting the free exciton peaks X$^0$. The EL trion peaks are asymmetric (unlike the PL trion), so asymmetric pseudo-Voigt functions\cite{Stancik08} (yellow solid lines) were used instead. The physical origin of the long-tail on the low-energy side of the trion peak is most likely inelastic trion-electron scattering,\cite{Manassen96} considering the trions form in an electron-rich environment. Below each sample label, the p-n junction gap between topgates is stated.}
    \label{FigS-ELfits}
\end{figure}

\clearpage

\section{EL lineshape fits and peak identification}
\label{sec:ELfits}

\begin{figure}[t]
    \includegraphics[width=1.0\columnwidth]{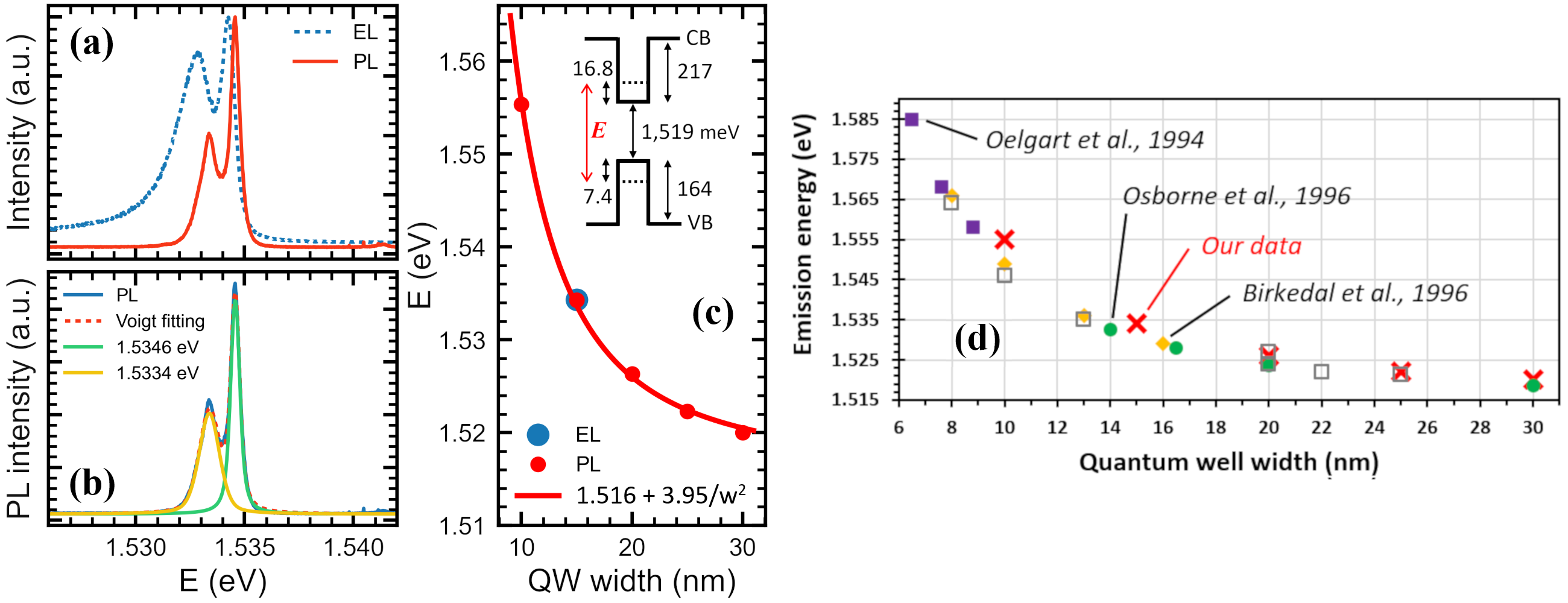}
    \caption{\textbf{(a)} Comparison of PL and EL spectra from sample B at $T=3$ K.   \textbf{(b)} Peak fitting of the two PL peaks shown in panel (a). The solid blue line is the experimental data. The dashed red line is the integrated fit, including contribution from both peaks. The green solid line is the fit to X$^0$, and the yellow solid line is the fit to X$^-$. \textbf{(c)(main)} PL emission energies of X$^0$ from a series of six nominally identical dopant-free GaAs quantum wells heterostructures except for a varying width $w$, ranging from 10 nm to 30 nm (red circles). \textbf{(c)(inset)} Energies in meV (not to scale) used to calculate $E_{\text{bX}}$ in Eqn.\,(S1). \textbf{(d)} The PL X$^0$ emission energies from our six quantum well heterostructures (of which two have nominally identical quantum well widths) are plotted (red crosses) alongside values taken from literature for PL X$^0$ in GaAs quantum well widths ranging from 6 nm to 30 nm. Series of quantum well widths (three or more) from a single publication are shown in color: closed purple squares (Oelgart \textit{et al}.\cite{Oelgart94}), closed orange diamonds (Birkedal \textit{et al}.\cite{Birkedal96}), and closed green circles (Osborne \textit{et al}.\cite{OsborneJL96}). Various data for a single value of quantum well width from a single publication are indicated by open grey squares.\cite{Shields94,FinkelsteinG95,KumarR96,Manassen96,Hayne99,Esser00}}
    \label{Fig-PLfits}
\end{figure}

There are two EL peaks at $\sim$\,1.533~eV and $\sim$\,1.534~eV in each of the four panels of Figure \ref{FigS-ELfits}. Neither of these two peaks correspond to light emission from the semi-insulating (SI) GaAs substrate ($\sim$\,1.519~eV), nor from the MBE growth buffer's superlattice (see Figure \ref{Fig-MBE-Fab}a) containing twenty GaAs quantum wells 2.5 nm wide (roughly corresponding to emission at $\sim$\,1.64 eV), nor from any of the well-known donor/acceptor impurities ($E<1.516$~eV) in GaAs.\cite{Bogardus68,Pavesi94} The temperature dependence of both peaks (no saturation from 85 K down to 3 K), partially shown in Figure \ref{FigS-Tdep}a, also rules out emission from impurities/traps. The temperature dependence also revealed another peak, whose energy difference from the $\sim$\,1.534~eV peak is consistent with the energy difference between heavy and light holes (see Figures \ref{FigS-Tdep} and \ref{FigS-LHcompare}). Lastly, the EL emission energy of the $\sim$\,1.534~eV peak is consistent with the PL emission energies of free excitons as a function of quantum well width in the literature (see Figure \ref{Fig-PLfits}d). We thus attribute the narrowest peak ($E \sim 1.534$ eV) to the neutral exciton X$^0$ ground state for a 15 nm wide GaAs quantum well. We identify the lower-energy peak ($E \sim 1.533$ eV) to charged excitons (trions), due to its characteristic temperature dependence [see Figure \ref{FigS-Tdep}a], its lower energy and wider linewidth than X$^0$. Its charge is likely to be negative (X$^-$), since there are more electrons than holes in the quantum well for the same topgate voltage magnitude (see plot of carrier density versus topgate voltage in Figure \ref{FigS-transport}), and most of the EL emanated from underneath the thin semi-transparent Ti topgate on the 2DEG side.

\section{Non-resonant photoluminescence of G375 and G569}
\label{sec:PLfits}

Figure \ref{Fig-PLfits}a compares G375's non-resonant PL spectrum and the EL spectrum from sample B; the comparison in Figure 3b of the main text was between PL from G375 and EL from sample C. Figure \ref{Fig-PLfits}b shows the lineshape fits of the two PL peaks, the neutral exciton X$^0$ and negatively-charged trion X$^-$, with symmetric Voigt functions. The emission energies for X$^0$ match very well, with $E_{\textsc{el}}=1,534.4$~meV and $E_{\textsc{pl}}=1,534.6$~meV (same $E_{\textsc{pl}}$ for both wafers G375 and G569). The FWHM of the PL X$^0$ peak is 0.51~meV from the lineshape fit. It is smaller than any of the FWHM for EL X$^0$ listed in Table I from the main text, because EL occurs at much higher carrier densities (causing more electron-exciton scattering), higher electric fields (from both $V_{\text{pn}}$ and $V_{\text{tg}}$), and possibly higher temperatures (due to Joule heating from the p-n current). On one hand,  a similar trend is observed for the PL and EL X$^-$ trions, where the FWHM of EL X$^-$ is much larger than that of PL X$^-$, presumably for the same reasons as EL X$^0$. On the other hand, the emission energies for the PL and EL trions X$^-$ do not match, the lineshape of EL X$^-$ is asymmetric whereas it is symmetric for PL X$^-$, and EL X$^-$ occurs at a much lower energy than PL X$^-$. The latter implies the binding (dissociation) energy of EL X$^-$ is larger than that of PL X$^-$. These differences between EL X$^-$ and PL X$^-$ are the same for all samples A$-$E and for both wafers G375 and G569.

The two heterostructures used in the main text (G375 and G569) both had a quantum well width of 15 nm, and have identical PL emission energies. Both wafers were part of a series of six nominally identical dopant-free GaAs quantum wells heterostructures except for a varying width $w$, ranging from 10 nm to 30 nm. Figure \ref{Fig-PLfits}c shows the PL emission energies (red circles) for each wafer in the series. The PL X$^0$ energies scale as $1/w^2$ (best fit represented by solid red line), as expected for bound states in a square quantum well. The PL emission energies of X$^-$ also scale as $1/w^2$ with a slightly larger prefactor (not shown), which is consistent with the higher optical mass of charged excitons/trions. Figure \ref{Fig-PLfits}d compares the PL X$^0$ emission energies with those from the literature: overall, our PL data is consistent with literature.

 \begin{figure}[t]
    \includegraphics[width=0.60\columnwidth]{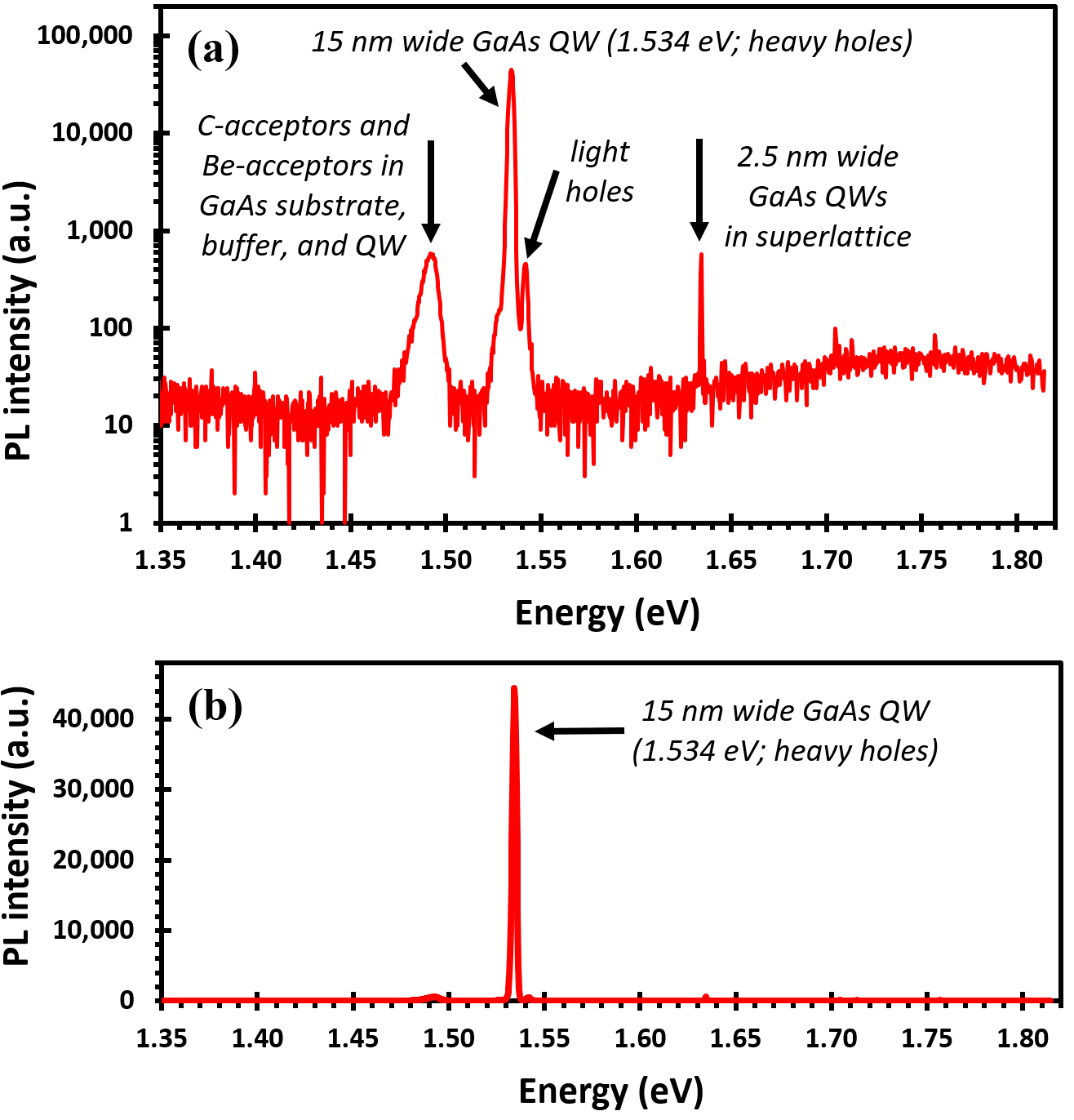}
    \caption{Wide-energy PL spectrum of quantum well (QW) heterostructure G375. (a) Data shown on a log scale. (b) Same data, shown on a linear scale. No cleanroom processing was performed on the wafer sample shown here. Unlike all other data in this paper, the data shown above was acquired with a low-resolution grating of 400 lines/mm.}
    \label{FigS-fullPL}
\end{figure}

Figure \ref{FigS-fullPL} shows a wide PL spectrum of wafer G375, from 1.35 eV (918 nm) to 1.82 eV (681 nm). The main peak at $\sim\,$1.534 eV is in reality the merger of two peaks: one peak from free neutral excitons involving heavy holes (labeled X$^0$ in the main text) and the other from charged excitons (trions, labeled X$^-$ in the main text). The peak on the high-energy side of the main peak, near $\sim\,$1.541 eV, is from neutral free excitons involving light holes (labeled LH in the main text). The emission line near $\sim\,$1.634 eV is from the superlattice (SL) of twenty thin (2.5 nm wide) GaAs quantum wells in the MBE growth buffer. Finally, the very broad peak near $\sim\,$1.493 eV comes from C acceptors throughout the heterostructure (including the substrate) and Be acceptors in the layers grown by MBE (the chamber contains a Be cell). Overall, the PL spectrum is very clean, with the main peak two orders of magnitude brighter than the next brightest peak. A high-resolution scan of the dominant peak near $\sim\,$1.534 eV can be found in Figure \ref{Fig-PLfits}b, where the PL X$^0$ and X$^-$ peaks can be fully resolved.

\section{Binding energies of EL X$^0$ and EL X$^-$}
\label{sec:E_bX}

Table I from the main text reveals that, as the p-n junction gap between the $V_{\text{tgL}}$ and $V_{\text{tgR}}$ topgates decreases from 2000 nm to 200 nm, both X$^-$ and X$^0$ show a very weak Stark shift (0.6~meV) to lower EL energies due to the increasing in-plane electric field $|\vec{E}_{\text{ext}}| = V_{\text{pn}}$/gap, ranging from 7.8 kV/cm (in sample E) to 78 kV/cm (in sample A) in the 2DEG/2DHG plane. With out-of-plane electric fields (of order 10 kV/cm), a large decrease (of order $\sim$20 meV) in $E_{\textsc{pl}}$ has been reported with photoluminescence.\cite{Miller85,Brum85,HongS87,Sanders87,Nojima88,Chuu91} However, with in-plane electric fields, the only effects reported\cite{Miller85} are an absence of any energy shift in $E_{\textsc{pl}}$ and a rapid broadening of PL peaks, leading to the near-complete smearing of all PL peaks by 16 kV/cm. Unlike past PL experiments with in-plane electric fields, we do not observe EL linewidths to rapidly increase with increasing in-plane electric fields, and all our EL linewidths are much narrower (0.7$-$0.9 meV from Table I in main text) than the reported PL linewidths\cite{Miller85} (9.4$-$10.0 meV) at their narrowest. Our narrow EL linewidths in turn allowed the observation of the small but noticeable shift\footnote{The decrease in $E_{\textsc{el}}(\text{X}^0,\text{X}^-)$ is not due to Joule heating (which would cause the bandgap to decrease): sample A, which has the smallest p-n junction gap and hence the smallest electrical resistance across the gap (i.e.~the smallest heat dissipation), has the largest decrease in $E_{\textsc{el}}$, whereas the converse is true for sample E (i.e. the largest p-n junction gap but the smallest decrease in $E_{\textsc{el}}$).} in $E_{\textsc{el}}$; this small shift would likely have been masked by the rapid and large linewidth broadening reported in the original PL experiments.

A possible reason for the near-absence of broadening is that $V_{\text{pn}} = 1.55$~V was kept constant for all our samples, such that the in-plane electric field was only changed by the size of the gap between the $V_{\text{tgL}}$ and $V_{\text{tgR}}$ topgates. This limited the amount of voltage broadening\cite{Bagwell89-Tavg} in EL spectra. By contrast,
to achieve an in-plane electric field of 16 kV/cm in Ref.~\onlinecite{Miller85}, a source-drain bias of 40 V (with a 2.7 \textmu A current) was applied across the 25 \textmu m gap (with 15 M\textohm~resistance) between two ohmic contacts. Such a large bias window (26 times larger than $V_{\text{pn}} = 1.55$~V) would have caused significant voltage broadening\footnote{The concept of voltage broadening did not gain wide acceptance among experimentalists until the late 1990s, now colloquially known as ``electron temperature'' (the combination of thermal and voltage broadening) when 2DEGs are involved.} to PL spectra in the original experiment.

The emission energy of EL X$^0$ can be decomposed into the following contributions:
\begin{equation}
E_{\textsc{el}}(\text{X}^0)=E_{\text{g}} - \vec{p} \cdot \vec{E}_{\text{ext}} + E_{\text{e-}\textsc{qw}} + E_{\text{h-}\textsc{qw}} - E_{\text{b}\textsc{x}}
\label{eq:EbX}
\end{equation}
\noindent where $E_{\text{g}}$ is the GaAs bandgap energy, $\vec{p}$ is the free exciton dipole moment, $\vec{E}_{\text{ext}}$ is the in-plane electric field, $E_{\text{e-}\textsc{qw}}$ ($E_{\text{h-}\textsc{qw}}$) is the lowest-energy bound state in the quantum well conduction (valence) band CB (VB), and $E_{\text{b}\textsc{x}}$ is the binding energy of the neutral exciton. Energies used in our calculations are shown in the inset of Figure~\ref{Fig-PLfits}c: the GaAs bandgap $E_{\text{g}}=1,519.0$~meV, the conduction band (CB) offset 217.0~meV, the valence band (VB) offset 164.0~meV, the quantum well bound state energies for electrons ($E_{\text{e-}\textsc{qw}}=16.8$~meV) and for holes ($E_{\text{h-}\textsc{qw}}=7.4$~meV), and the emission energy (red; PL or EL). We used 1,519.0~meV and 1,900.0~meV for the bandgaps of GaAs\cite{Pavesi94} and Al$_{0.3}$Ga$_{0.7}$As\cite{Lourenco01} respectively. The conduction band offset at the GaAs/AlGaAs interface is assumed to be $\sim 0.6 \Delta E_g$.\cite{Batey86,WangGuo13} The quantum well bound state energies are numerically found  from the standard textbook finite barrier square well transcendental equation. We use the isotropic mass $m^*_e=0.067\,m_0$ for electrons in our 15 nm wide quantum well, where $m_0$ is the electron rest mass $9.11 \times 10^{-31}$~kg. Heavy holes have different masses in different crystal directions.\cite{Filinov04} We use an average effective hole mass, the optical mass\cite{Pavesi94} $m^*_{hh}=(m^*_z\,m^*_y\,m^*_x)^{1/3}$, where $m^*_z=0.377\,m_0$ in the [001] crystal direction (MBE growth direction) and $m^*_y=m^*_x=0.112\,m_0$ in the [110] and [1$\bar{1}$0] crystal directions [eqn.~(24) in Ref.~\onlinecite{Winkler95}]. Thus, taking into account the small redshift $\vec{p} \cdot \vec{E}_{\text{ext}}$ due to the Franz-Keldysh effect,\cite{Franz58,Keldysh58} the spatial asymmetry in the effective mass of heavy holes, the bandgaps of GaAs and Al$_{0.3}$Ga$_{0.7}$As and their band offsets,\cite{Pavesi94, Lourenco01,Batey86,WangGuo13} and the energy of the ground state for holes and electrons in the GaAs quantum well, the binding energy of EL X$^0$ is calculated to be $E_{\text{b}\textsc{x}} = (8.8\pm0.2)$~meV (see Table I in the main text). The same calculations and results apply to PL X$^0$ from wafers G375 and G569.

\begin{figure}[h]
    \includegraphics[width=0.75\columnwidth]{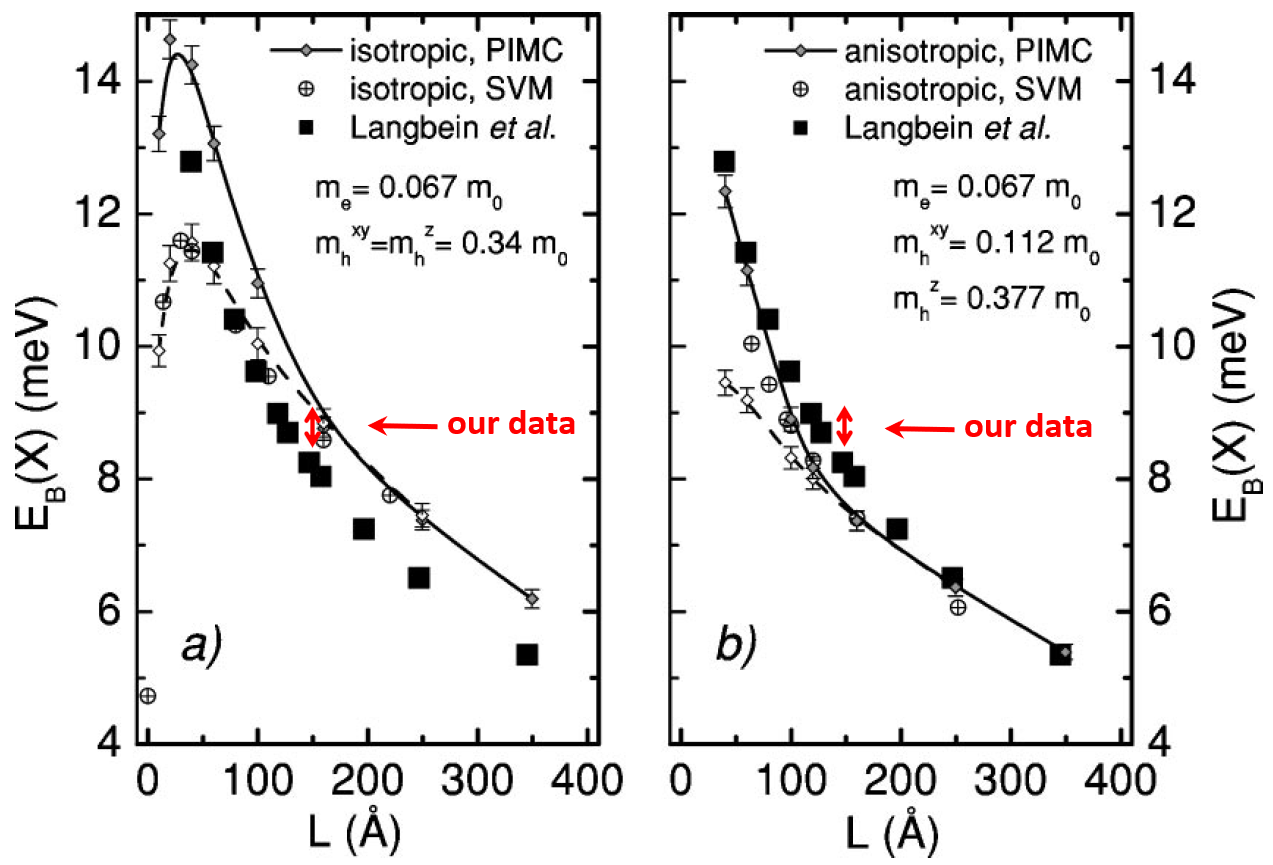}
    \caption{Comparison of our EL exciton binding energy with theory. The EL X$^0$ energies range from 1,534.1 to 1,534.7 meV (see Table I in the main text). Using those values and Eqn.~(S1), we find that $E_{\text{b}\textsc{x}} = (8.8\pm0.3)$~meV for samples A$-$E. Our result (with its uncertainty represented by the red double-ended vertical arrow) is compared to experiment\cite{Langbein99-B} (filled black squares) and theory\cite{Filinov04} (circles and lines) in Figure 10 from Ref.~\onlinecite{Filinov04} (shown above), where theoretical calculations used (a) isotropic and (b) anisotropic hole effective masses. [Figure adapted from Ref.~\onlinecite{Filinov04}]}
    \label{Fig:EbX}
\end{figure}

Figure \ref{Fig:EbX} shows our $E_{\text{b}\textsc{x}}$ values for PL/EL X$^0$ are consistent with reported experiments\cite{Oelgart94,Birkedal96,Langbein99-B} and theory\cite{Oelgart94,Filinov04,Winkler95,AndreaniLC90-A} on PL X$^0$. Our experimental values are slightly larger than the experimental PL $E_{\text{b}\textsc{x}}$ values from Ref.~\onlinecite{Langbein99-B}, but fall well within the maximum/minimum theoretical values (solid black lines in panels a and b). The maximum/minimum theoretical values are dictated by the choice of isotropic/anisotropic hole effective mass symmetry.

Unexpectedly, the trion dissociation energy appears to increase with higher in-plane electric field (see Table I). There is no ambiguity as to whether $\Delta E_{\text{X}^-}$ is increasing or not with in-plane electric field, since it is directly obtained from the difference in emission energies $\Delta E_{\text{X}^-} = E_{\textsc{el}}(\text{X}^0) - E_{\textsc{el}}(\text{X}^-)$. The emission energy of EL X$^-$ can be decomposed into the following contributions:
\begin{equation}
E_{\textsc{el}}(\text{X}^-) = E_{\text{g}} - \vec{p} \cdot \vec{E}_{\text{ext}} + E_{\text{e-}\textsc{qw}} + E_{\text{h-}\textsc{qw}} - E_{\text{b}\textsc{x}} - \Delta E_{\text{X}^-}.
\label{eq:DeltaEX}
\end{equation}
\noindent An increase in the dissociation energy of trions with larger in-plane electric field has been experimentally observed and theoretically justified in InGaAs/GaAs quantum dots,\cite{Saito15} in the presence of strong spin-orbit coupling.

\begin{figure}[t]
    \includegraphics[width=0.80\columnwidth]{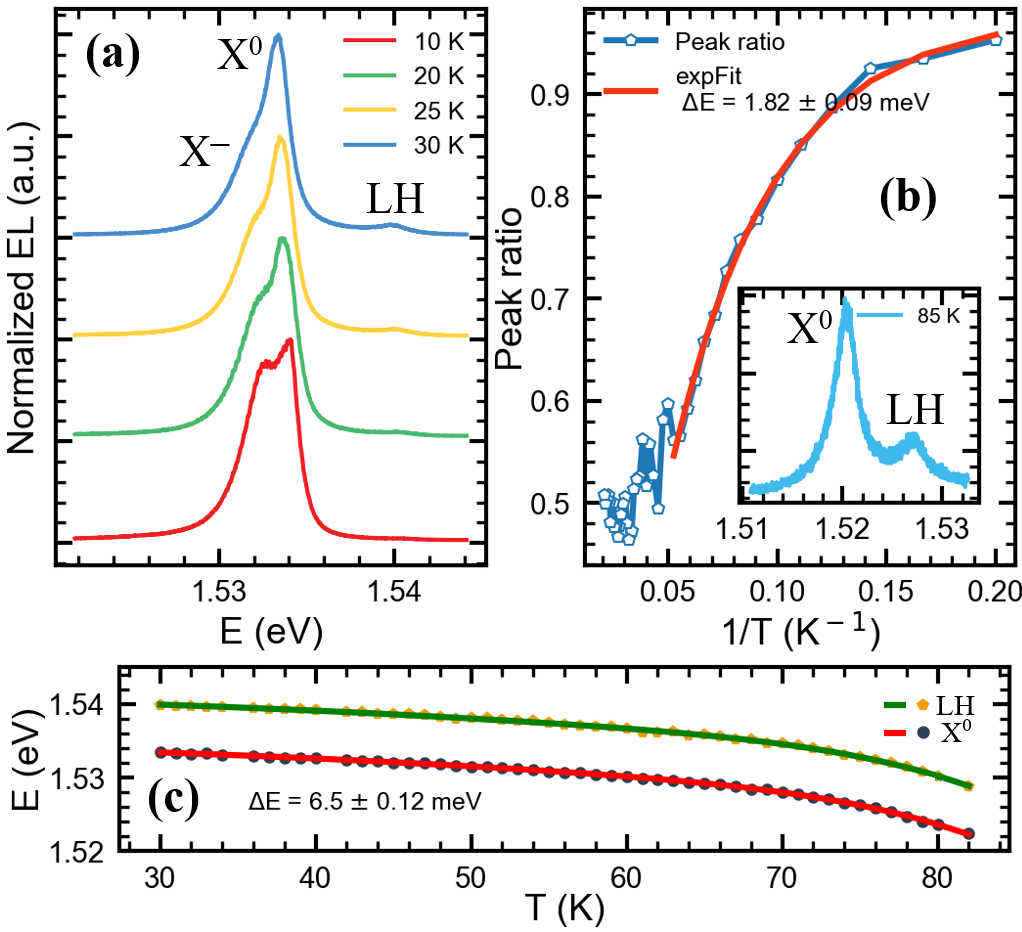}
    \caption{Temperature dependence of EL in sample~A. (a) Temperature dependence of EL from sample A operated at Set-Reset frequency of $f_{\textsc{sr}}=1$~kHz. The four spectra shown have been normalized so that the X$^0$ peak height is constant. The spectra have been vertically offset for clarity. (b)(main) Fit of the peak intensity ratio X$^-$/X$^0$ as a function of temperature. (b)(inset) EL intensity as a function of energy in eV at 85 K: the X$^0$ peak can still be resolved. The LH peak is evident at higher temperatures, but not at lower temperatures. (c) EL energies of X$^0$ and LH as a function of temperature, from experiments (circles) and theory (lines). The latter uses P{$\ddot{\mbox{a}}$}ssler's semi-empirical model\cite{Passler01,Passler02} for the GaAs bandgap.}
    \label{FigS-Tdep}
\end{figure}

\section{Detailed temperature dependence of EL}
\label{sec:Tdep}

Figure \ref{FigS-Tdep}a shows the temperature dependence of the X$^0$ and X$^-$ EL peaks. As expected for a charged exciton, X$^-$ disappears above a much lower temperatures threshold ($T > 30$~K) than X$^0$. The X$^0$ peak survives up to $T \approx 85$~K, as shown in the inset of Figure \ref{FigS-Tdep}b. The PL energy separation between X$^0$ and X$^-$ in Figure \ref{Fig-PLfits}b suggests a dissociation energy of 1.2 meV for the trion's 2nd electron, consistent with similarly obtained energies with PL in GaAs quantum wells.\cite{Shields95-A} The EL energy separation between X$^0$ and X$^-$ for all samples in Figure~\ref{FigS-ELfits} (see also Table I in main text) show a range of dissociation energies spanning 1.4$-$1.8 meV for the trion's 2nd electron, for the same quantum well width. Another experimental method for measuring the dissociation energy involves fitting the ratio of integrated peak intensity between X$^0$ and X$^-$ with $e^{-\Delta E/kT}$ for a wide range of temperatures.\cite{Platt11} Using this method for the EL spectra, Figure \ref{FigS-Tdep}b yields (1.8$\pm$0.1) meV for the trion's dissociation energy, consistent with the 1.8 meV value obtained from Figure \ref{FigS-ELfits}a at $T=3$~K in sample A. For both X$^0$ and X$^-$, the EL intensity decreases as the temperature increases, due to some combination of exciton dissociation, activation of non-radiative recombination centers, and thermally-induced escape of carriers from the quantum well. From $T=3$~K to $T=85$~K, the emission energy of both X$^0$ and LH redshifts by $\sim$\,14~meV, mostly due to the temperature-dependent decrease of the GaAs bandgap.\cite{Passler01,Passler02}

\section{Identifying light holes in electroluminescence}
\label{sec:LH}

\begin{figure}[t]
    \includegraphics[width=0.70\columnwidth]{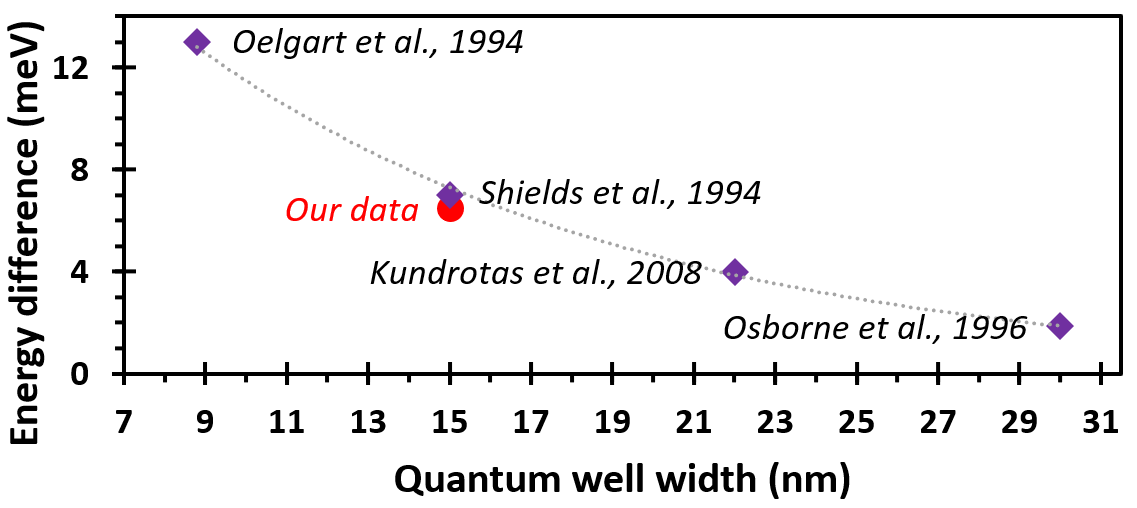}
    \caption{Comparison with literature of the difference in emission energies between heavy/light hole free excitons. Literature values for PL (purple diamonds) come from Refs.\,\onlinecite{Shields94,Oelgart94,OsborneJL96,Kundrotas08}. The dotted line is a guide to the eye. Our EL data (red circles) is consistent with the trend.}
    \label{FigS-LHcompare}
\end{figure}

A third EL peak, labeled ``LH'', is visible in Figure \ref{FigS-Tdep}, appearing $\sim$\,6.5 meV above X$^0$ for a wide range of temperatures. It also appears at the same emission energy in the PL spectra of Figure \ref{Fig-PLfits}a (barely visible, just above $E \sim 1.541$~eV) and of Figure \ref{FigS-fullPL}a, also 6.5 meV above X$^0$. Excitons composed of an electron bound to a light hole can co-exist with excitons composed of an electron bound to a heavy hole, and have higher PL emission energies than the latter due to their smaller effective mass ($m^*_{lh}=0.082m_0$).\cite{Pavesi94} Characteristically, the ratio of the PL intensity of the light-hole free exciton to that of the heavy-hole free exciton increases with temperature, due to the thermally-driven transfer of holes from the heavy-hole band to the light-hole band.\cite{Shields95-B,DangGT09,OsborneJL96} This behavior is indeed observed in the EL shown in Figure \ref{FigS-Tdep}a, and continues until the highest temperature attempted ($T=86$ K) in the inset of Figure \ref{FigS-Tdep}b. The 6.5 meV energy difference between EL X$^0$ and LH (see Figure \ref{FigS-Tdep}c) for our 15 nm quantum well width is consistent with those observed with PL in modulation-doped quantum wells of various widths (see Figure~\ref{FigS-LHcompare}).\cite{Shields94,Oelgart94,OsborneJL96,Kundrotas08} This 6.5 meV difference is also consistent with the calculated energy difference ($\approx6.4$~meV) between the heavy-hole and light-hole square-well bound states in the valence band (see inset of Figure \ref{Fig-PLfits}c). We therefore identify the LH peak with the ground state of light-hole free excitons. This peak also appears in other dopant-free devices (but is not identified as such): see EL spectrum in Figure 4b of Ref.~\onlinecite{ChungYC19}, the peak nearest wavelength 806 nm.

\section{Transport characterization, mobility and density}
\label{sec:transport}

\begin{figure}[t]
    \includegraphics[width=1.0\columnwidth]{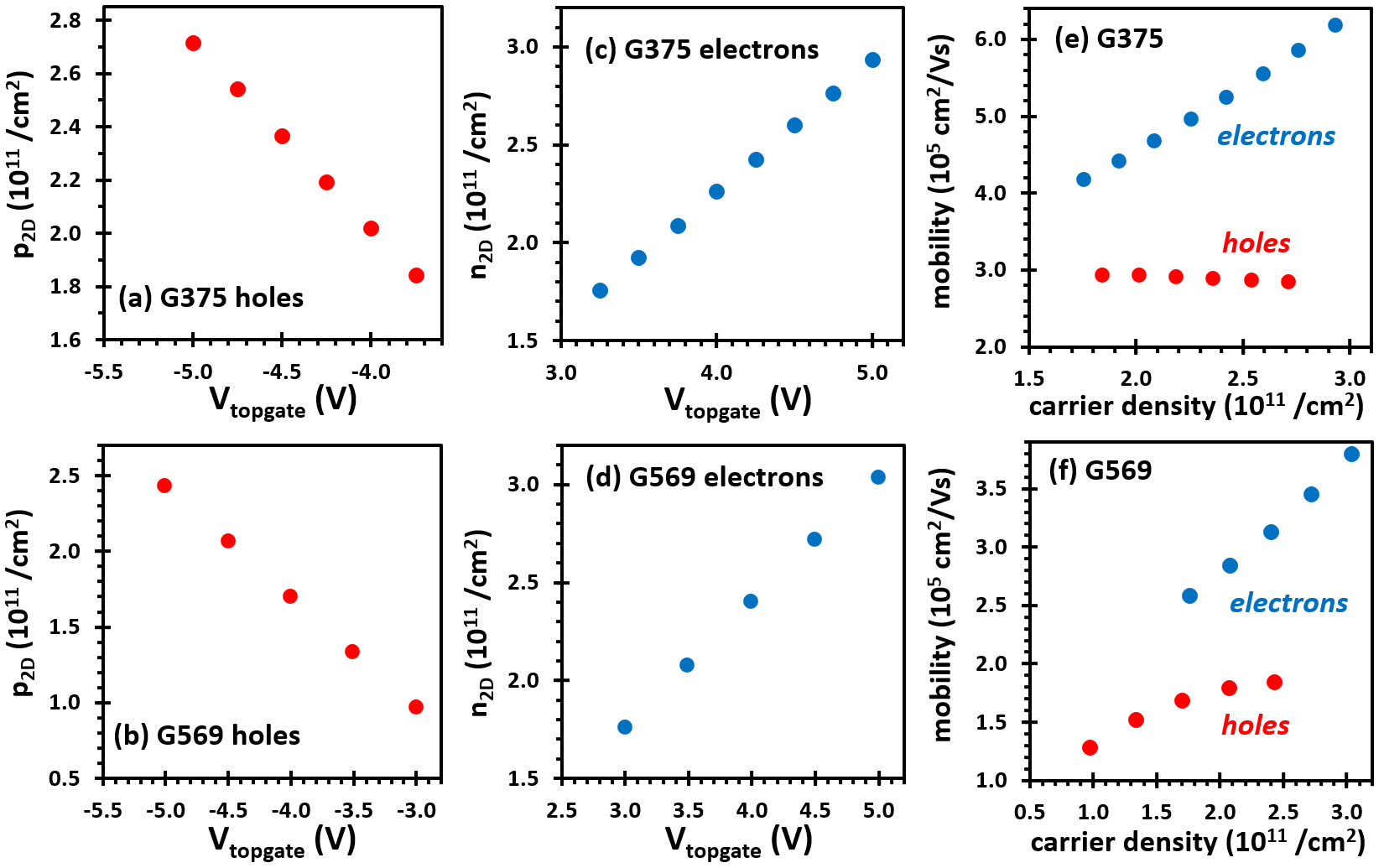}
    \caption{All four-terminal measurements shown above were taken at $T=1.5$\,K on dedicated  gated Hall bars. Hole density ($p_{\textsc{2d}}$) versus topgate voltage for wafer: (a) G375 and (b) G569. Electron density ($n_{\textsc{2d}}$) versus topgate voltage for wafer: (c) G375 and (d) G569. Electron and hole mobilities versus carrier density for wafer: (e) G375 and (f) G569.}
    \label{FigS-transport}
\end{figure}

Figure \ref{FigS-transport} shows the gating and mobility characteristics of wafers G375 and G569. In all cases, the carrier density versus topgate voltage relation is linear and reproducible, with no time drift or voltage hysteresis. In both wafers, the electron mobilities are relatively low compared to maximum values achieved in dopant-free GaAs 2DEGs,\cite{Arjun22,Yousun19,Croxall13} because their quantum well width is too arrow (only 15 nm) to maximize electron mobility.\cite{Croxall13} Hole mobilities are consistent with those achieved in other dopant-free narrow quantum well heterostructures.\cite{Yousun19,Croxall13} In both G375 and G569, the electron mobility exceeds the hole mobility because of the difference in effective mass between electrons ($m^* = 0.067 m_e$) and holes ($m^* = 0.35 m_e$) where $m_e$ is the electron rest mass. Because of their smaller size, holes are in general more susceptible than electrons to the GaAs/AlGaAs interface roughness scattering mechanism, and thus hole mobility saturates at much lower carrier density than it does in electrons. In G375, the mobility has already peaked for the density used in the experiments described in the main text, whereas it is just reaching its peak in wafer G569.

\section{Self-consistent band structure simulations}
\label{sec:nextnano}

\begin{figure}[t]
    \includegraphics[width=0.75\columnwidth]{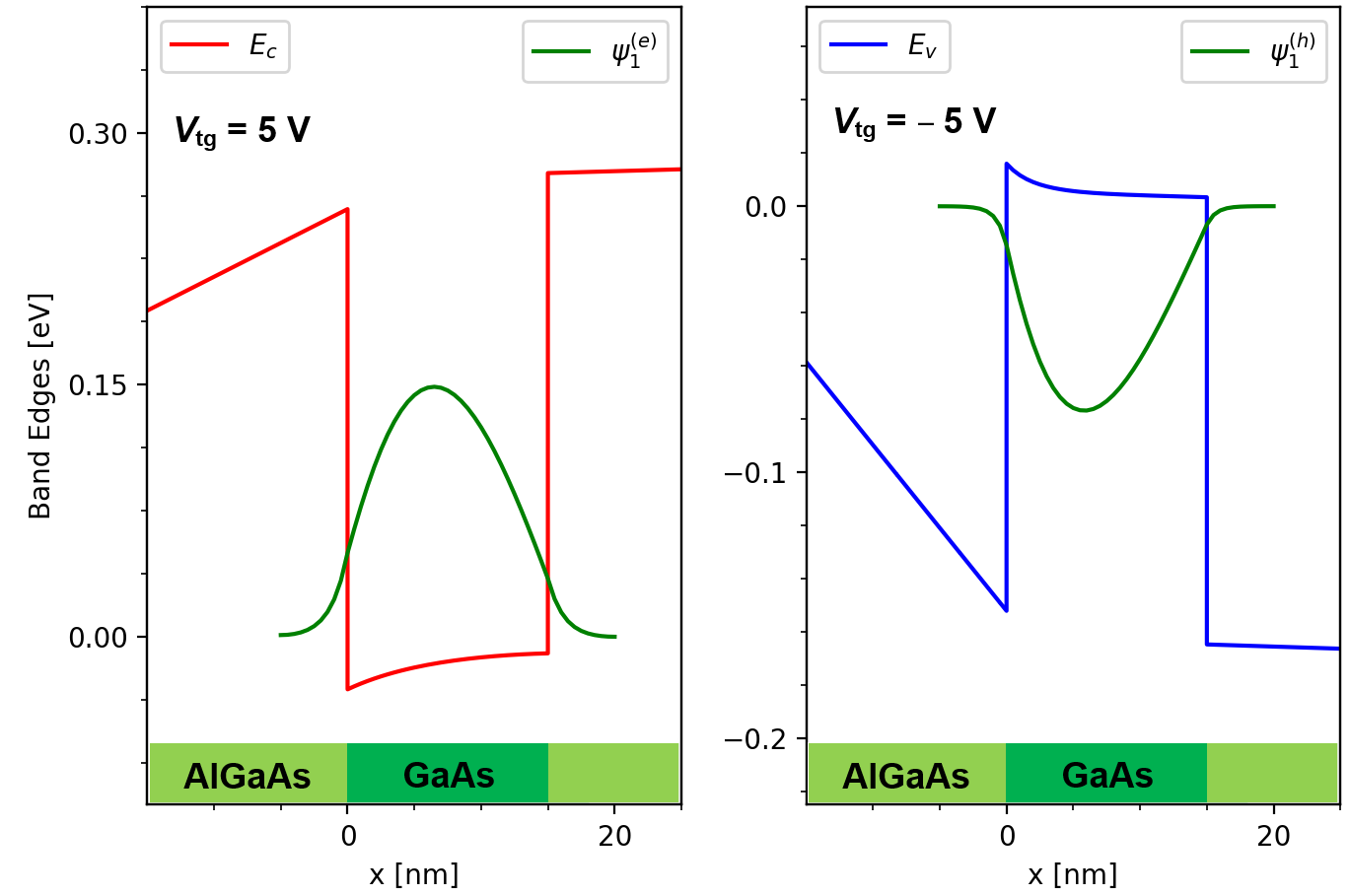}
    \caption{Band structure simulations of an ambipolar dopant-free quantum well heterostructure for a: (a) 2DEG wavefunction with $V_{\text{tg}}=+5$ V, and (b) 2DHG wavefunction with $V_{\text{tg}}=-5$ V. The GaAs/AlGaAs quantum well is represented at the bottom of the panels. $E_{\textsc{c}}$ is the conduction band edge, $E_{\textsc{v}}$ is the valence band edge, and $V_{\text{tg}}$ is the topgate voltage.}
    \label{FigS-bandstructure-z}
\end{figure}

Figure \ref{FigS-bandstructure-z} shows the calculated band structures and wavefunction profiles in the MBE growth direction of a dopant-free GaAs/AlGaAs quantum well, using nextnano$^{\tiny{\text{\textregistered}}}$ software.\cite{NextNano-A,NextNano-B,NextNano-C}
As-grown, the wafer is not conductive: a gated field-effect transistor (FET) must be fabricated. As seen in panels (a)$-$(d) in Figure \ref{FigS-transport}, such a FET does not conduct at $V_{\text{tg}}=0$ V, because the band structure is essentially flat. At $V_{\text{tg}}=+5$ V ($-5$ V), the lowest 2D subband of the GaAs quantum well is populated and a 2DEG (2DHG) has formed.

\clearpage

